\pdfoutput=1

\documentclass[11pt,twoside,a4paper,cmspaper,final,collab]{cms-tdr}
\def\svnVersion{680260b-D}\def\svnDate{2020/01/30}\def\cmsCernNoTag{CERN-EP-2019-164}\def\cmsCernDate{\today}\def\cmsMessage{"Published in the Journal of High Energy Physics as \href{http://dx.doi.org/10.1007/JHEP01(2020)096}{\doi{10.1007/JHEP01(2020)096}.}"}

\begin{document}\cmsNoteHeader{HIG-18-004}

\hyphenation{had-ron-i-za-tion}
\hyphenation{cal-or-i-me-ter}
\hyphenation{de-vices}
\RCS$HeadURL$
\RCS$Id$

\newcolumntype{x}[1]{D{.}{.}{#1}}
\newcommand{\mc}[1]{\multicolumn{1}{c}{#1}}
\providecommand{\NA}{\ensuremath{\text{---}}}
\newlength\cmsTabSkip\setlength{\cmsTabSkip}{1ex}
\newlength\cmsTabSkipLarge\setlength{\cmsTabSkipLarge}{5ex}
\providecommand{\cmsTable}[1]{\resizebox{\textwidth}{!}{#1}}
\ifthenelse{\boolean{cms@external}}{\providecommand{\cmsTabSpace}{\hspace{-2ex}}}{\providecommand{\cmsTabSpace}{}}

\newlength\cmsFigWidth
\ifthenelse{\boolean{cms@external}}{\setlength\cmsFigWidth{0.49\columnwidth}}{\setlength\cmsFigWidth{0.49\textwidth}}
\ifthenelse{\boolean{cms@external}}{\providecommand{\cmsLeft    }{left\xspace}}{\providecommand{\cmsLeft}{left\xspace}}
\ifthenelse{\boolean{cms@external}}{\providecommand{\cmsRight}{right\xspace}}{\providecommand{\cmsRight}{right\xspace}}
\ifthenelse{\boolean{cms@external}}{\providecommand{\cmsBottom}{bottom\xspace}}{\providecommand{\cmsBottom}{bottom\xspace}}
\ifthenelse{\boolean{cms@external}}{\providecommand{\cmsTop}{top\xspace}}{\providecommand{\cmsTop}{top\xspace}}

\providecommand{\CL}{CL\xspace}

\newcommand{\thelumi}{\ensuremath{35.9\fbinv}\xspace}
\newcommand{\sqrtsEight}{\ensuremath{\sqrt{s}=8\TeV}\xspace}
\newcommand{\sqrtsThirteen}{\ensuremath{\sqrt{s}=13\TeV}\xspace}
\newcommand{\tanbeta}{\ensuremath{\tanb}\xspace}
\newcommand{\PHp}{\ensuremath{\PH^+}\xspace}
\newcommand{\mHpm}{\ensuremath{m_{\PH^\pm}}\xspace}
\newcommand{\mh}{\ensuremath{m_{\Ph}}\xspace}
\newcommand{\mA}{\ensuremath{m_{\PSA}}\xspace}
\newcommand{\Hptbprod}{\ensuremath{\cPaqt\cPqb\PH^{+}+\cPaqt\PH^{+}}\xspace}
\newcommand{\Hptballprod}{\ensuremath{\cPqt(\cPqb)\PH^{+}}\xspace}
\newcommand{\tbHp}{\ensuremath{\cPqt\to\PH^{+}\cPqb}\xspace}
\newcommand{\mhmodm}{\ensuremath{m_\mathrm{h}^\mathrm{mod-}}\xspace}
\newcommand{\mhonetwentyfive}{\ensuremath{M_\mathrm{h}^\mathrm{125}(\tilde{\chi})}\xspace}
\newcommand{\ttbb}{$\cPqt\cPaqt$+$\cPqb$($\cPqb$)}
\newcommand{\ttcc}{$\cPqt\cPaqt$+$\cPqc$($\cPqc$)}
\newcommand{\ttlf}{$\cPqt\cPaqt$+LF}
\newcommand{\Nbjets}{\ensuremath{N_{\cPqb\,\text{jets}}}}
\newcommand{\Njets}{\ensuremath{N_\text{jets}}}
\newcommand{\BHptb}{\ensuremath{\mathcal{B}(\PH^+\!\to\!\cPqt\cPaqb)}}
\newcommand{\sPPtoHptb}{\ensuremath{\sigma(\Pp\Pp\!\to\PH^{+}\cPaqt\cPqb)}}
\newcommand{\sPPtoHpBoth}{\ensuremath{\sigma(\Pp\Pp\!\to\PH^{+}\cPaqt\cPqb + \Pp\Pp\!\to\PH^{+}\cPaqt)}}
\newcommand{\BHmtb}{\ensuremath{\mathcal{B}(\PH^-\!\to\!\cPaqt\cPqb)}}
\newcommand{\sPPtoHmtb}{\ensuremath{\sigma(\Pp\Pp\!\to\PH^{-}\cPqt\cPaqb)}}
\newcommand{\sPPtoHmBoth}{\ensuremath{\sigma(\Pp\Pp\!\to\PH^{-}\cPqt\cPaqb + \Pp\Pp\!\to\PH^{-}\cPqt)}}
\newcommand{\sPPtoHtb}{\ensuremath{\sigma_{\PH^\pm}}}
\newcommand{\BHtb}{\ensuremath{\mathcal{B}(\PH^\pm\!\to\!\cPqt\cPqb)}}
\providecommand{\cmsTable}[1]{\resizebox{\textwidth}{!}{#1}}

\cmsNoteHeader{HIG-18-004}
\title{Search for a charged Higgs boson decaying into top and bottom quarks in events with electrons or muons in proton-proton collisions at $\sqrt{s}=13\TeV$ }

\date{\today}

\abstract{
A search is presented for a charged Higgs boson heavier than the top quark, produced in association with a top quark, or with a top and a bottom quark, and decaying into a top-bottom quark-antiquark pair. The search is performed using proton-proton collision data collected by the CMS experiment at the LHC at a center-of-mass energy of 13\TeV, corresponding to an integrated luminosity of 35.9\fbinv. Events are selected by the presence of a single isolated charged lepton (electron or muon) or an opposite-sign dilepton (electron or muon) pair, categorized according to the jet multiplicity and the number of jets identified as originating from \cPqb{} quarks. Multivariate analysis techniques are used to enhance the discrimination between signal and background in each category. The data are compatible with the standard model, and 95\% confidence level upper limits of 9.6--0.01\unit{pb} are set on the charged Higgs boson production cross section times branching fraction to a top-bottom quark-antiquark pair, for charged Higgs boson mass hypotheses ranging from 200\GeV to 3\TeV. The upper limits are interpreted in different minimal supersymmetric extensions of the standard model.
}

\hypersetup{
pdfauthor={CMS Collaboration},
pdftitle={Search for a charged Higgs boson decaying into top and bottom quarks  in events with electrons or muons in proton-proton collisions at 13 TeV},
pdfsubject={CMS},
pdfkeywords={CMS,  dnn, machine learning, data driven, 2hdm, charged higgs, 13 TeV, MSSM, hmssm, mhmodp, mhmodm}}

\maketitle

\section{Introduction}
\label{sec:introduction}

Since the discovery of a Higgs boson~\cite{Aad:2012tfa,Chatrchyan:2012xdj,Chatrchyan:2013lba} with a mass of 125\GeV~\cite{Aad:2015zhl,Sirunyan:2017exp},
the ATLAS and CMS Collaborations have actively searched for additional neutral and charged Higgs bosons.
Most theories beyond the standard model (SM) of particle physics
enrich the SM Higgs sector; a simple extension is the assumption of the existence of two Higgs doublets~\cite{Gunion:2002zf,Akeroyd:2016ymd,Branco:2011iw,Gildener:1976ih}.
Such models are collectively labeled as \textit{two-Higgs-doublet models} (2HDM),
and are further classified into four categories according to the couplings of the doublets to fermions.
In Type-I models, only one doublet couples to fermions,
while in Type-II models one doublet couples to the up-type quarks and the other to the down-type quarks and the charged leptons.
In lepton-specific models one doublet couples only to the leptonic sector and the other couples to quarks,
while in flipped models the first doublet couples specifically to the down-type quarks and the second one to the up-type quarks and charged leptons.

{
The two-doublet structure of the 2HDM Higgs sector gives rise to five physical Higgs bosons through spontaneous symmetry breaking:
a charged pair (\PHpm) and three neutral bosons, namely the light (\Ph) and heavy (\PH) scalar Higgs bosons, and one pseudoscalar boson (\PSA).
Supersymmetric (SUSY) models have a Higgs sector based on 2HDMs~\cite{Fayet:1974pd,Fayet:1976et,Fayet:1977yc,Dimopoulos:1981zb,Sakai:1981gr,Inoue:1982ej}.
Among the SUSY models, a popular one is the minimal supersymmetric extension to the SM (MSSM)~\cite{Djouadi:2005gj,Carena:2013ytb},
whose Higgs sector is described by a Type-II 2HDM.
In the MSSM, the production and decay of these particles are described at tree level by two free parameters,
which can be chosen as the mass of the charged Higgs boson (\mHpm) and the ratio of the vacuum expectation values of the neutral components of the two Higgs doublets
(\tanbeta).\par}

Some variants of the 2HDM achieve consistency with the 125\GeV Higgs boson via a Gildener-Weinberg scalon scenario which stabilizes the Higgs boson mass and alignment~\cite{PhysRevD.99.055015}.

\begin{figure}[tb]
\centering\includegraphics[width=.8\textwidth]{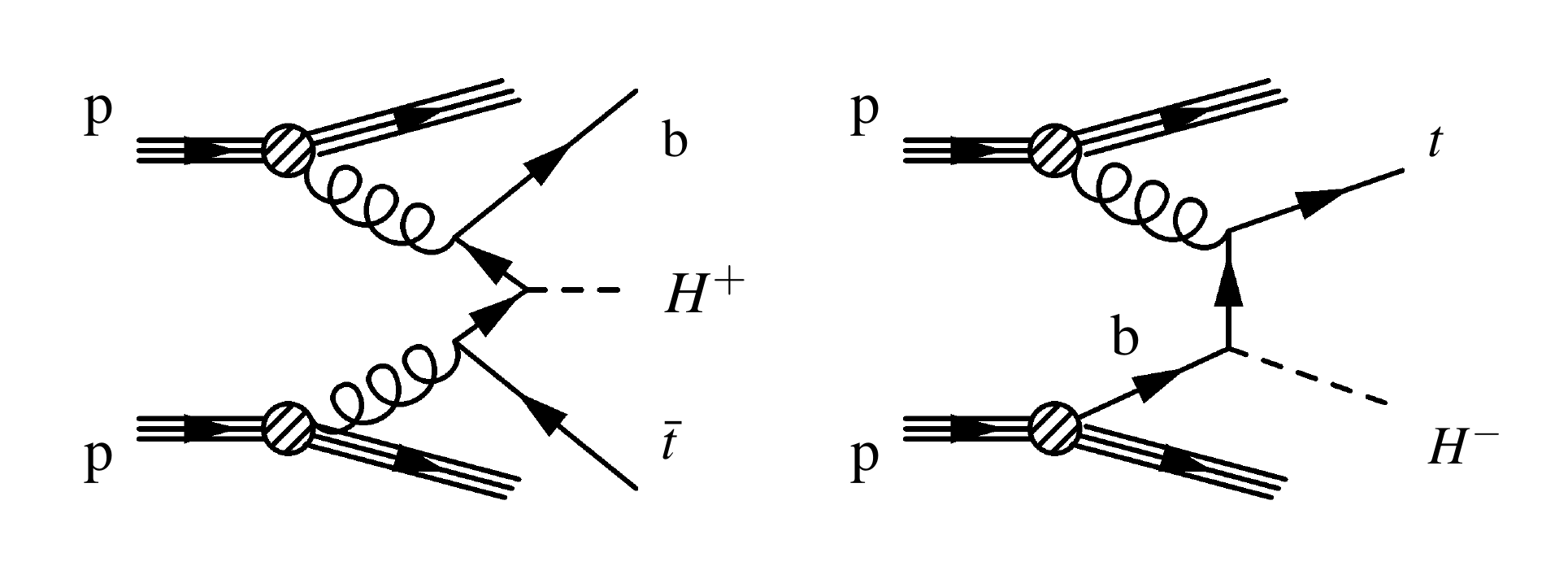}
\caption{\label{fig:diagram} Feynman diagrams for the production of a heavy charged Higgs boson in the four-flavor scheme (4FS, left)
and in the five-flavor scheme (5FS, right).}
\end{figure}

Charged Higgs bosons with a mass below the top quark mass are dominantly produced in top quark decays,
whereas charged Higgs bosons with a mass larger than the top quark mass are produced in association with a top quark.
Charged Higgs boson production at finite order in perturbation theory is accomplished
in association with a top and a bottom quark in the so-called \textit{four-flavor scheme} (4FS)
and in association with a top quark
in the \textit{five-flavor scheme} (5FS)~\cite{Harlander:2011aa}, as illustrated in Fig.~\ref{fig:diagram}.

In this paper,
only charged Higgs bosons with a mass larger than the mass of the top quark (\textit{heavy charged Higgs bosons}) are considered,
and charge-conjugate processes are implied. The signal is produced in the 4FS, and the eventual presence of a 5FS production is accounted for in the search region definition. The normalization of the signal processes accounts for both the 4FS and the 5FS.

The decay of a heavy charged Higgs boson can occur through several channels,
among them $\PHp\to\Pgt^{+}\Pgngt$ and $\PHp\to\cPqt\cPaqb$
have the highest branching fractions, respectively at low (about 200\GeV) and high (about 1\TeV) \mHpm for a large range of \tanbeta values and a large variety of theoretical models~\cite{deFlorian:2016spz}.

The detection of a charged Higgs boson would unequivocally point to physics beyond the SM.
Model-independent searches for charged Higgs bosons are of utmost interest for the CERN LHC program because they allow one to disentangle the Higgs sector physics from the specificity and complexity of the theoretical model by assuming unity branching fraction in each mode.

Direct searches for charged Higgs bosons have been performed by the CERN LEP and the Fermilab Tevatron experiments, and indirect constraints on \Hpm production have been set from flavor physics measurements~\cite{20021,Abdallah:2003wd,Abreu:1999bx,Achard:2003gt,Abbiendi:2008aa,Abbiendi:1998rd,Abbiendi:2013hk,Aaltonen:2009vf,Abazov:2011up,Arbey:2017gmh}.
Searches for a charged Higgs boson decaying into a top and a bottom quark have been performed by the D0, ATLAS, and CMS Collaborations
in proton-antiproton collisions at a center-of-mass energy of $\sqrt{s}=1.96$\TeV~\cite{Abazov:2008rn} and in proton-proton ($\Pp\Pp$) collisions at \sqrtsEight~\cite{Khachatryan:2015qxa,Aad:2015typ} and \sqrtsThirteen~\cite{Aaboud:2018cwk}. In this paper we improve the sensitivity to model-independent production of a charged Higgs boson, as well as the sensitivity to relevant MSSM scenarios.
The ATLAS and CMS Collaborations have also conducted searches for the production of a charged Higgs boson in the $\Pgt^{+}\Pgngt$~\cite{Aaboud:2018gjj,Khachatryan:2015qxa,Aaboud:2016dig,Aad:2014kga}, \ensuremath{\mathrm{\PQc\overline{\PQs}}}~\cite{Khachatryan:2015uua}, and \ensuremath{\mathrm{\PQc\overline{\PQb}}}~\cite{Sirunyan:2018dvm} decay channels at $\sqrt{s} = 8$ and 13\TeV.

Searches for charged Higgs bosons produced via vector boson
fusion and decaying into \PW\ and \PZ\ bosons, as predicted by models containing Higgs triplets~\cite{Sirunyan:2017sbn,PhysRevLett.114.231801,Sirunyan:2018taj}, and
searches for additional neutral heavy Higgs bosons decaying to a pair of third-generation fermions \ttbar{}, $\bbbar{}$, and \ensuremath{\Pgt^{+}\Pgt^{-}}~\cite{Sirunyan:2018taj,Sirunyan:2018zut,Aaboud:2017sjh,Aaboud:2017hnm,Sirunyan:2017uyt}
extend the program of the ATLAS and CMS Collaborations to elucidate the extended Higgs sector beyond the SM.

This paper describes a search for a heavy charged Higgs boson produced in association with a top quark or with a top and a bottom quark and decaying into a top and a bottom quark performed using $\Pp\Pp$ collision data collected at \sqrtsThirteen in 2016. The data correspond to an integrated luminosity of \thelumi.
The final state contains two \PW~bosons, one from the decay chain of the heavy charged Higgs boson and the other from the decay of the associated top quark. One or both of the \PW~bosons can decay into leptons, producing single-lepton and dilepton final states, respectively. The leptonic decays of tau leptons from the \PW~boson decay are considered as well.
The single-lepton final state is characterized
by the presence of one isolated lepton (\Pe, \Pgm) that is used to trigger the event, while the dilepton final state contains events with two isolated opposite-sign leptons ($\Pep\Pem$, $\Pe^\pm\Pgm^\mp$, $\Pgmp\Pgmm$).
This leads to the suppression of several backgrounds.
The signal process (\Hptbprod) has furthermore a large \cPqb~jet multiplicity;
an additional classification of the events is therefore achieved based on the number of jets identified as originating from \cPqb~quarks.

Multivariate analysis (MVA) techniques are used to enhance the discrimination between signal and background.
Signal-rich regions are analyzed together with signal-depleted regions in a maximum likelihood fit to the MVA classifier outputs,
which simultaneously determines the contributions from the \Hptbprod signal and the backgrounds.

Model-independent upper limits
on the product of the charged Higgs boson production cross section and the branching fraction into
a top-bottom quark-antiquark pair, $\sPPtoHtb\BHtb=\sPPtoHpBoth\BHptb+\sPPtoHmBoth\BHmtb$,
as a function of \mHpm, are presented in this paper.
Results are also interpreted
in specific MSSM benchmark scenarios,
where many free parameters of the model are fixed to values corresponding to interesting phenomenological assumptions.

\section{The CMS detector}
\label{sec:detector}

The central feature of the CMS apparatus is a superconducting solenoid of 6\unit{m} internal diameter, providing a magnetic field of 3.8\unit{T}. Within the solenoid volume are a silicon pixel and strip tracker, a lead tungstate crystal electromagnetic calorimeter (ECAL), and a brass and scintillator hadron calorimeter (HCAL), each composed by a barrel and two endcap sections. Forward calorimeters extend the pseudorapidity ($\eta$) coverage provided by the barrel and endcap detectors. Muons are detected in gas-ionization chambers embedded in the steel flux-return yoke outside the solenoid.
Events of interest are selected using a two-tiered trigger system~\cite{Khachatryan:2016bia}.
The first level, composed by specialized hardware processors, uses information from the calorimeters and muon detectors,
while the second level consists of a farm of  processors running a version of the full event reconstruction software optimized for fast processing.
A more detailed description of the CMS detector, together with a definition of the coordinate system used and the relevant kinematic variables, can be found in Ref.~\cite{Chatrchyan:2008zzk}.

\section{Event simulation}
\label{sec:datasets}

Signal events are simulated using the \MGvATNLO~2.3.3~\cite{Alwall:2014hca} generator at next-to-leading order (NLO) precision in perturbative quantum chromodynamics (QCD) using the 4FS for a range of \mHpm hypotheses between 200 and 3000\GeV; the complete list of masses is $[200, 220, 250, 300, 350, 400, 500, 650, 800, 1000, 1500, 2000, 2500, 3000]$\GeV. The 4FS is expected to provide a better description of the observables, while shape effects from 5FS production are expected to be negligible, because eventual additional \cPqb{} quarks would be radiated with low transverse momentum by the beam remnants~\cite{deFlorian:2016spz}.

Normalization effects induced by the presence of 5FS are accounted for by computing the MSSM production cross sections for the heavy charged Higgs boson signals both in the 4FS and 5FS; the two cross sections are then combined to obtain the total cross section using the \textit{Santander matching} scheme \cite{Harlander:2011aa}
for different values of \tanbeta.
The 4FS and 5FS cross sections differ for all mass point by about 20\%, and the Santander-matched cross section lies inbetween the two; typical values are of the order of 1\unit{pb} for a mass of 200\GeV, down to about $10^{-4}$\unit{pb} for a mass of 3\TeV~\cite{Heinemeyer:1998yj,deFlorian:2016spz,Berger:2003sm,Flechl:2014wfa,Degrande:2015vpa,Dittmaier:2009np}.

Branching fractions $\BHptb$ are computed in the chosen scenarios with the \HDECAY~6.52 package~\cite{Djouadi:1997yw}. These cross sections are used in Section~\ref{sec:results} only for the model-dependent results, and don't affect the model-independent results.

The main background to this analysis originates from SM top quark pair production.
Other backgrounds are the production of \PW{} and $\PZ/\Pgg^{\scriptscriptstyle *}$ with additional jets (referred to as V+jets), diboson and triboson processes, single top quark production, \ttbar production in association with \PW{}, \PZ{}, \Pgg{}, or \PH~bosons (collectively labeled \ttbar{}+V), as well as four top quark production (\ttbar{}\ttbar{}) and QCD multijet events.

The \ttbar{}, $\ttbar\PH$, and single top quark events in the $t$- and $\cPqt\PW$-channels are generated at NLO precision in perturbative QCD with \POWHEG v2.0 \cite{Frixione:2007vw,Nason:2004rx,Alioli:2010xd}.

The \MGvATNLO~2.2.2 generator~\cite{Alwall:2014hca} is used at leading order (LO), with the MLM jet matching and merging~\cite{Alwall:2007fs}, to generate vector boson events in association with jets, single top quark events in the $s$-channel, and four top quark production.
The associated production of \ttbar events with a vector boson and with a \Pgg is simulated at NLO using \MGvATNLO~2.2.2 with FxFx jet matching and merging~\cite{Frederix:2012ps}.

{\tolerance=800
In all cases, the NNPDF3.0~\cite{Ball:2014uwa} set of parton distribution functions (PDFs) is used, and
the parton showers and hadronization processes are performed by \PYTHIA8.212~\cite{Sjostrand:2014zea}
with the CUETP8M1~\cite{Skands:2014pea} tune for the underlying event,
except for the \ttbar sample where the tune CUETP8M2T4~\cite{Khachatryan:2015pea}
provides a more accurate description of the kinematic distributions of the top quarks and of the jet multiplicity.\par}

Next-to-NLO (NNLO) calculations are used to compute the cross section for the dominant \ttbar background
for a top quark mass of 172.5\GeV,
including resummation to next-to-next-to-leading-logarithmic
accuracy ~\cite{Cacciari:2011hy,Baernreuther:2012ws,Czakon:2012zr,Czakon:2012pz,Beneke:2011mq,Czakon:2013goa,Czakon:2011xx}.
The other backgrounds are normalized using NLO (single top quark $t$- and $s$-channels~\cite{Aliev:2010zk,Kant:2014oha}, $\ttbar$+V production~\cite{Maltoni:2015ena}, and diboson production~\cite{Campbell:2011bn}), NNLO (V+jets production),
and approximate NNLO (single top quark $\cPqt\PW$-channel~\cite{Kidonakis:2010ux}) cross sections.

The simulated \ttbar events are further separated based on the flavor
of additional jets that do not originate from the top quark decays in the event
and are labeled according to their content in \cPqb{}- and \cPqc{}-originated hadrons.
The \ttbb{}  (\ttcc{}) label is attributed
to the events that have at least one \cPqb~jet (\cPqc~jet and no \cPqb~jet) from the event generator within the acceptance.
Events that do not belong to any of the above processes are enriched in light-flavor jets and therefore denominated as \ttlf.
This partition of the simulated $\ttbar$ sample is based on matching heavy-flavor generator-level jets to the originating partons and hadrons and is introduced to account for different systematic uncertainties affecting the corresponding cross section predictions. The procedure is detailed in Refs.~\cite{Bartosik:2047049,Cacciari:2007fd}.

All generated events are passed through a detailed simulation of the CMS apparatus, based on \GEANTfour~v9.4~\cite{Agostinelli:2002hh}.
The effects of additional $\Pp\Pp$ interactions occurring in the same or in neighboring bunch crossings (pileup) are modelled by adding
simulated minimum bias events to all simulated processes.
In the data collected in 2016 an average of 23 $\Pp\Pp$ interactions occurred per LHC bunch crossing.
In simulation, the difference in the number of true interactions is accounted for by reweighting the simulated events to match the data in the multiplicity distribution of pileup interactions.

\section{Event reconstruction}
\label{sec:obj}

Events are reconstructed using the particle-flow (PF) algorithm~\cite{Sirunyan:2017ulk}, which aims to reconstruct and identify each individual particle in an event, with an optimized combination of information from the various elements of the CMS detector. The energy of photons is obtained from the ECAL measurement. The energy of electrons is determined from a combination of the electron momentum at the primary interaction vertex as determined by the tracker, the energy of the corresponding ECAL cluster, and the energy sum of all bremsstrahlung photons spatially compatible with originating from the electron track. The momentum of muons is obtained from the curvature of the corresponding track. The energy of charged hadrons is determined from a combination of their momentum measured in the tracker and the matching ECAL and HCAL energy deposits, corrected for zero-suppression effects and for the response function of the calorimeters to hadronic showers. Finally, the energy of neutral hadrons is obtained from the corresponding corrected ECAL and HCAL energy.
The reconstructed vertex with the largest value of summed physics-object squared transverse momentum ($\pt^2$) is taken to be the primary $\Pp\Pp$ interaction vertex~\cite{Chatrchyan:2014fea}.
The physics objects are the jets, clustered using the jet finding algorithm~\cite{Cacciari:2008gp,Cacciari:2011ma} with the tracks assigned to the vertex as inputs, and the associated missing transverse momentum (\ptvecmiss), taken as the negative vector sum of the \pt of those jets.

Electrons are identified using an MVA-based identification algorithm~\cite{Khachatryan:2015hwa}.
Working points are defined~\cite{Sirunyan:2018omb} by setting thresholds for the classifier values to mitigate efficiency losses for high-\pt electrons observed particularly in high-mass signal events;
such working points are labeled \textit{Tight} ($\approx$88\% efficiency for \ttbar events)
and \textit{Loose} ($\approx$95\% efficiency for \ttbar events).
They result in an efficiency in selecting high-mass signal events of $\approx$90\%, approximately flat across the electron high-\pt range.
Muon identification uses the algorithm described in Ref.~\cite{Sirunyan:2018fpa} and two working points, referred to as \textit{Medium} and \textit{Loose}, with efficiencies of about 97 and 100\%, respectively.
Thresholds in \pt~and $\eta$ for electrons and muons depend on whether they are used for selecting or vetoing events and are detailed in Section~\ref{sec:Overview}.

Electrons and muons are required to be isolated from other particles. Their relative isolation is measured as
the ratio between the scalar \pt sum of selected PF particles within a cone of a radius $\Delta R(\pt(\ell))$ and the \pt of the particle; $\Delta R$ is defined as $\sqrt{\smash[b]{(\Delta\eta)^2+(\Delta\phi)^2}}$
and $\Delta \eta$ and $\Delta \phi$ are the distances in the pseudorapidity and azimuthal angle.
The $\Delta R(\pt(\ell))$ cone decreases with the lepton \pt~\cite{Rehermann:2010vq,Khachatryan:2016kod} according to the formula
\begin{equation}
  \Delta R(\pt(\ell)) = \frac{10\GeV}{\min\big[\max(\pt(\ell),50\GeV), 200\GeV\big]}.
\end{equation}
Efficiencies in triggering, reconstruction, identification, and isolation of leptons are estimated both in data and simulation.
Those efficiencies are used to determine correction factors, depending on $\pt$ and $\eta$, and are applied to simulated events on a per-lepton basis.

Jets are reconstructed from the PF particles clustered by the anti-\kt algorithm~\cite{Cacciari:2008gp,Cacciari:2011ma} with a clustering radius of 0.4.
To mitigate the effect of pileup interactions, charged hadrons that do not arise from the primary vertex are excluded from the clustering.
Furthermore, jets originating from pileup interactions are removed by means of an MVA identification algorithm~\cite{CMS-PAS-JME-16-003}.
The jet momentum is then corrected in simulated events to account for multiple effects, including the extra energy clustered in jets arising from pileup.
In situ measurements of the momentum balance in dijet, photon+jet, \cPZ{}+jet, and multijet events are used
to determine any residual differences between the jet energy scale in data and in simulation,
and appropriate corrections are applied~\cite{Khachatryan:2016kdb}.
Jets are selected if they satisfy $\pt>40\GeV$ and $\abs{\eta}<2.4$. Loose identification criteria are applied to the jets, in order to distinguish them from well-identified stable particles.
Finally, jets are required to be separated from the selected leptons by $\Delta R>0.4$.

{\tolerance=800
Jets from the hadronization of \cPqb~quarks are identified ($\cPqb$ tagged) using the \textit{combined secondary vertex} algorithm~\cite{Sirunyan:2017ezt}.
For the chosen threshold of the tagging algorithm, the mistagging probability---the fraction of jets that arise from the fragmentation of light partons (\cPqu, \cPqd, \cPqs, and \cPg) and \cPqc~jets misidentified by the algorithm as \cPqb~jets---is approximately 1 and 15\%, respectively,
while the efficiency to correctly identify a \cPqb~jet is about 70\%.
The difference in \cPqb~tagging and mistagging efficiencies between data and simulation is corrected by applying correction factors dependent on jet \pt{} and $\eta$.\par}

The missing transverse momentum vector is defined as the projection of the negative vector sum of the momenta of all reconstructed PF particles
in an event onto the plane perpendicular to the beams. Its magnitude is referred to as \ptmiss.
The \ptvecmiss reconstruction is improved by propagating the effect of the jet energy corrections to it.
Further filtering algorithms are used to reject events with anomalously large \ptmiss resulting from instrumental effects~\cite{Khachatryan:2014gga}.

Hadronically decaying $\Pgt{}$ leptons ($\tauh$) are reconstructed using the \textit{hadron-plus-strips} algorithm~\cite{Sirunyan:2018pgf},
based on the identification of the individual $\Pgt$ decay modes.
The $\tauh$ candidates are required to be separated from reconstructed electrons and muons by $\Delta R > 0.4$.
Tau candidates are further selected by means of a multivariate discriminator combining isolation and lifetime information~\cite{Sirunyan:2018pgf}.
Jets originating from the hadronization of quarks and gluons misidentified as $\tauh$ are suppressed by requiring that the \tauh{} candidate
is isolated.
The $\tauh$ identification efficiency depends on $\pt^{\tauh}$ and $\eta^{\tauh}$, and is on average 50\%
for $\pt^{\tauh}>20\GeV$ with a probability of approximately 1\% for hadronic jets to be misidentified as a $\tauh$.
The isolation variable is constructed from the PF particles
inside a cone of $\Delta R=0.3$.
The effect of neutral PF candidates from pileup vertices is estimated using charged hadrons associated with those vertices and subtracted from the isolation variable.

\section{Event selection and classification}
\label{sec:Overview}

Events are selected with single-lepton triggers characterized by transverse momentum (\pt{}) thresholds of 27~(24)\GeV for electrons (muons).
Additionally,
several trigger paths with higher \pt{} thresholds and looser identification requirements
are included to maximize efficiency for high-\pt{} electrons (muons),
resulting in an overall efficiency in the plateau region close to 95 (100)\%.
Correction factors quantifying the difference between trigger efficiencies in data and simulated events are evaluated using a tag-and-probe technique~\cite{Chatrchyan:2012jra,Chatrchyan:2012xi,Khachatryan:2015hwa,Sirunyan:2018fpa}.

Events are required to have at least one electron (muon) with $\pt>35$ (30)\GeV
satisfying
tighter identification and isolation criteria than the online requirements, effectively corresponding to the saturation point of the online trigger efficiencies.
As briefly discussed in Section~\ref{sec:introduction}, the first classification is achieved by separating the events in five single-lepton and dilepton regions ($\Pe^\pm$, $\Pgm^\pm$, $\Pep\Pem$, $\Pe^\pm\Pgm^\mp$, $\Pgmp\Pgmm$).
In the single-lepton category, only events with exactly one lepton are accepted, whereas the presence of any additional lepton passing the loose identification requirements with $\pt>10\GeV$ vetoes the event.
Moreover, the presence of a $\tauh$ candidate with $\pt>20\GeV$ and $\abs{\eta}<2.3$ vetoes the event.
In the dilepton category, we accept events with exactly two oppositely charged leptons (electrons or muons); the second lepton is required to have $\pt>10\GeV$ and pass looser identification criteria than the leading lepton.
To reduce the $\PZ/\Pgg^{*}$ background, we reject events with two leptons of the same flavor and opposite charge with an invariant mass $m_{\ell\ell}$ less than 12 or between 76 and 106\GeV.

The final states examined in this paper include neutrinos
from the \PW~boson decays; events are therefore required to have $\ptmiss>30\GeV$.
Additionally, in the single-lepton final state, events in which the \ptmiss{} is compatible with mismeasurement of electron or jet energy are rejected by requiring the azimuthal angle separation between the \ptmiss{} and any jet in the event to be $\Delta\phi > 0.05$.

Tree-level signal production processes are characterized by having five (three) jets at leading order in the single-lepton (dilepton) final state.
The \ttbar background has a lower jet multiplicity in the corresponding regions, but
additional jets may be produced through initial- and final-state radiation.
Requiring a high multiplicity of reconstructed jets improves the discrimination of signal events from the background,
while the regions depleted in signal processes constrain background estimates using data.
Consequently, in the single-lepton and dilepton event regions, the presence of at least four and two jets, respectively, is required.
The SM top quark pair production
has final states similar to the charged Higgs boson signal production with fewer \cPqb~quarks at tree level, while additional gluon splitting contaminates the high \cPqb~jet multiplicity regions.
Consequently, one or more of these jets is required to be \cPqb-tagged.

Events are categorized according to the total number of associated jets $\Njets$ and
the \cPqb-tagged jet multiplicity $\Nbjets$,
yielding a total of nine regions in the single-lepton final state and eight regions in the dilepton final state.
In the single-lepton final state, the regions are: (4j/1\cPqb), (4j/2\cPqb), (4j/$\ge$3\cPqb), (5j/1\cPqb), (5j/2\cPqb), (5j/$\ge$3\cPqb), ($\ge$6j/1\cPqb), ($\ge$6j/2\cPqb), and ($\ge$6j/$\ge$3\cPqb);
while in the dilepton final states, where
less hadronic activity is expected,
the regions are: (2j/1\cPqb), (2j/2\cPqb), (3j/1\cPqb), (3j/2\cPqb), (3j/3\cPqb), ($\ge$4j/1\cPqb), ($\ge$4j/2\cPqb), and ($\ge$4j/$\ge$3\cPqb).
The resulting regions are characterized by different background compositions and signal purities, and are collectively labeled \textit{ signal regions} and used in the likelihood fit for signal extraction. We additionally define \textit{ control regions} which we use to correct from data the normalization of background samples; these regions are described in Section~\ref{sec:bkgSyst}.

For a large $\PH^{+}$ mass range, the highest significance for both the single-lepton and dilepton final states is found in the regions having higher $\Njets$ and $\Nbjets$.
The only exception are the $\PH^{+}$ signals with the mass around 200\GeV, where the low $\Njets$ and $\Nbjets$ regions have higher sensitivity than the high multiplicity ones.
Finally, events with two same-sign leptons are used to form control regions for the multijet background estimation.

A set of discriminant variables is selected
to enhance the signal and background separation in each category and is summarized in Table~\ref{tab:var}.

Kinematic and topological shapes have different discrimination power for the different mass hypotheses of the charged Higgs boson.
Each discriminant variable is studied and included in an MVA classifier if it improves the discrimination, or otherwise discarded.
For both single-lepton and dilepton regions,
the \HT{} distribution, defined as the scalar sum of the \pt{} of the selected jets, is one of the most sensitive variables.
Additionally, the largest \pt{} among the \cPqb~jets,
the \ptmiss,
the minimum invariant mass between the lepton and the \cPqb~jets, the maximum $\Delta\eta$ between two \cPqb-tagged jets,
the smallest $\Delta R$ separation of the \cPqb~jets, and
the \pt-weighted average of the \cPqb~tagging discriminator calculated using the non-\cPqb-tagged jets
are used as input variables to the MVA discriminators.
Information about the event topology is incorporated via event shape variables, such as
the centrality which is defined as the ratio of the sum of the transverse momenta of all jets to their total energy, and the second Fox--Wolfram moment~\cite{FoxWolfram} calculated using all jets.

In the single-lepton final states, the following variables are also included:
the invariant mass of the three jets with largest \pt{},
the transverse mass of the system constituted by the lepton and the \ptmiss,
the angular separation between the lepton and the system constituted by the \cPqb~jet pair with the smallest $\Delta R$ separation between the \cPqb~jets,
and the average separation between the \cPqb~jet pairs.

The event selection for the dilepton final state
takes advantage of the presence of the second lepton.
The lepton with largest \pt{} (leading lepton) characterizes the decay of a Lorentz-boosted top quark that originates from the massive charged Higgs boson in the signal hypothesis.
The following variables are also considered:
the $\Delta R$ between the leading lepton and the leading \cPqb-tagged jet,
the momentum of the leading lepton,
the lepton \pt{} asymmetry,
the mass of the lepton+\cPqb-tagged jet system with the largest \pt{}, and
the smallest of the transverse masses constructed with the leading \cPqb~jet and each of the two \PW~boson hypotheses, where the \PW~bosons are reconstructed using the \ptvecmiss and the lepton momenta.

\begin{table}
    \centering
    \topcaption{Summary of the discriminating variables used in the analysis of the single-lepton (1$\ell$) and dilepton (2$\ell$) final states.}
    \label{tab:var}
    \begin{tabular}{ccp{0.72\textwidth}}
        \multirow[c]{9}{*}{\rotatebox[origin=c]{90}{Common to $1\ell$ and $2\ell$}}
             & $H_\text{T}$ & Scalar sum of the jet transverse momenta\\
             & $\pt\null_{\cPqb}$ & Largest \pt{} among the \cPqb-tagged jets \\
             & $\ptmiss$ & Missing transverse momentum\\
             & $\min{m(\ell,\cPqb)}$ & Minimum invariant mass between the lepton and the \cPqb-tagged jet\\
             & $\max{\Delta\eta(\cPqb,\cPqb)}$ & Maximum pseudorapidity separation between \cPqb-tagged jet pairs\\
             & $\min{\Delta R (\cPqb,\cPqb)}$ & Minimum separation between \cPqb-tagged jet pairs \\
             & $\text{\pt-}\langle\text{CSV}\rangle$ & \pt weighted average of the combined secondary vertex discriminator of the non-\cPqb-tagged jets\\
             & $FW_2$ & Second Fox--Wolfram moment\\
             & centrality & Ratio of the sum of the \pt{} and the total energy of all jets\\[\cmsTabSkip]
        \hline
        \multirow{4}{*}{\rotatebox[origin=c]{90}{$1\ell$}}
             & $m_{jjj}$ & Invariant mass of the jet system composed by the first three jets ranked in \pt\\
             & $\mT(\ell,\ptvecmiss)$ & Transverse mass of the system constituted by the lepton and the \ptvecmiss\\
             & $\Delta R(\ell,\cPqb\cPqb)$ & Distance between the \cPqb-tagged jet pair with the smallest $\Delta R$ separation and the lepton \\
             & $\langle\Delta R(\cPqb,\cPqb)\rangle$ & Average separation between \cPqb-tagged jet pairs\\[\cmsTabSkip]
        \hline
        \multirow{7}{*}{\rotatebox[origin=c]{90}{$2\ell$}}
             & $\Njets$ & Number of selected jets \\
             & $\Nbjets$ & Number of selected \cPqb-tagged jets\\
             & $\Delta R(\ell,\cPqb)$ & Distance between the lepton and the \cPqb-tagged jet with largest transverse momenta\\
             & $\pt\null_\ell$ & Largest \pt{} between the leptons\\
             & $\frac{\pt\null_{\ell\,1}-\pt\null_{\ell\,2}}{\pt\null_{\ell\,1}+\pt\null_{\ell\,2}}$ & Lepton \pt asymmetry\\
             & $m(\ell,\cPqb)$ & Invariant mass of the lepton+\cPqb-tagged jet system with the largest \pt{} (top quark candidate)\\
             & $\mT^{min}$ & The smallest of the transverse masses constructed with the leading \cPqb-tagged jet and each of the two \PW~boson hypotheses: $\min{\left[\mT(\cPqb,\pt\null_{\ell\,1} +\ptvecmiss), \mT(\cPqb,\pt\null_{\ell\,2} +\ptvecmiss)\right]}$
        \\
    \end{tabular}
\end{table}

\begin{figure}[h!p]
\centering\includegraphics[width=\cmsFigWidth]{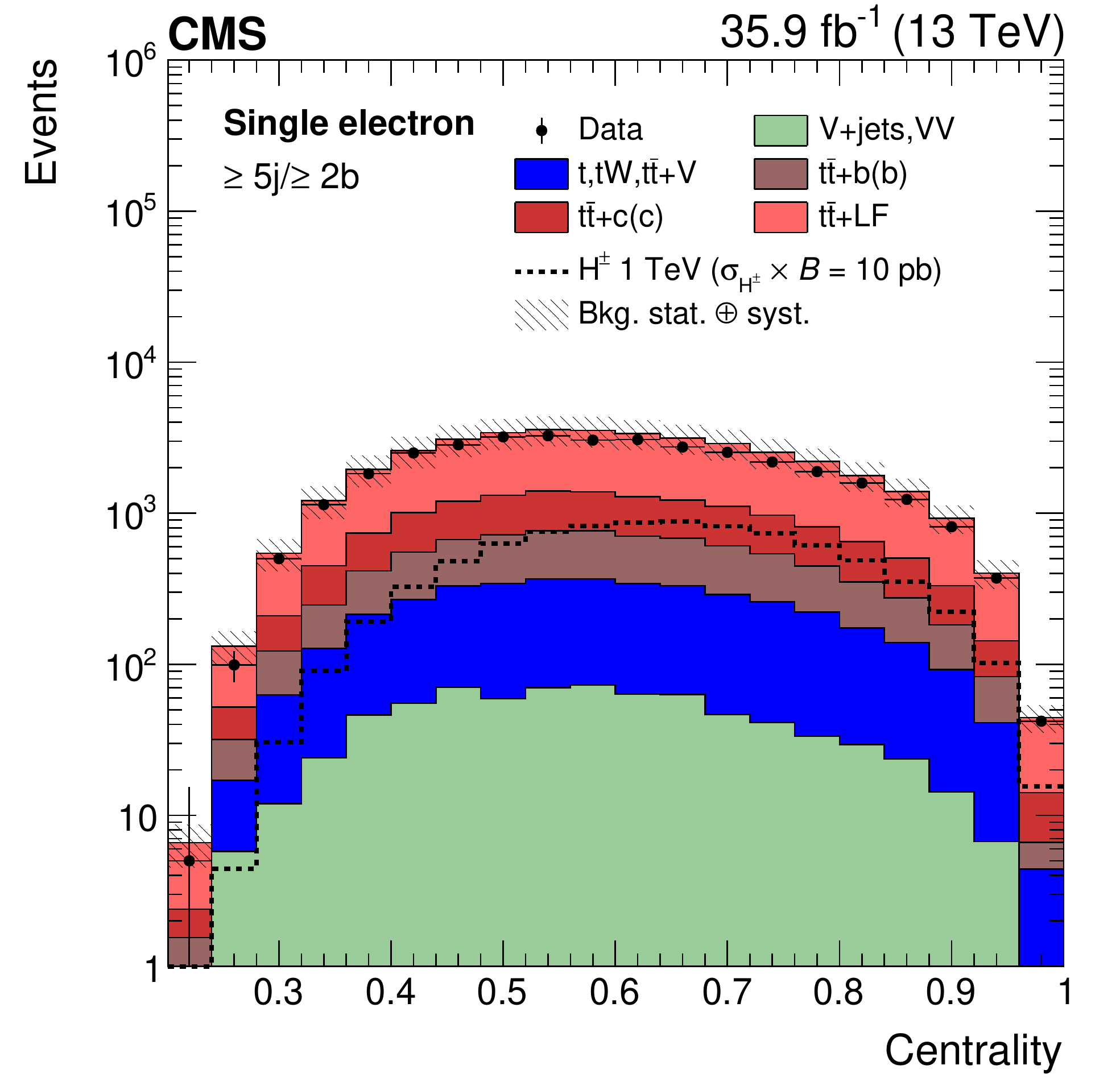}
\centering\includegraphics[width=\cmsFigWidth]{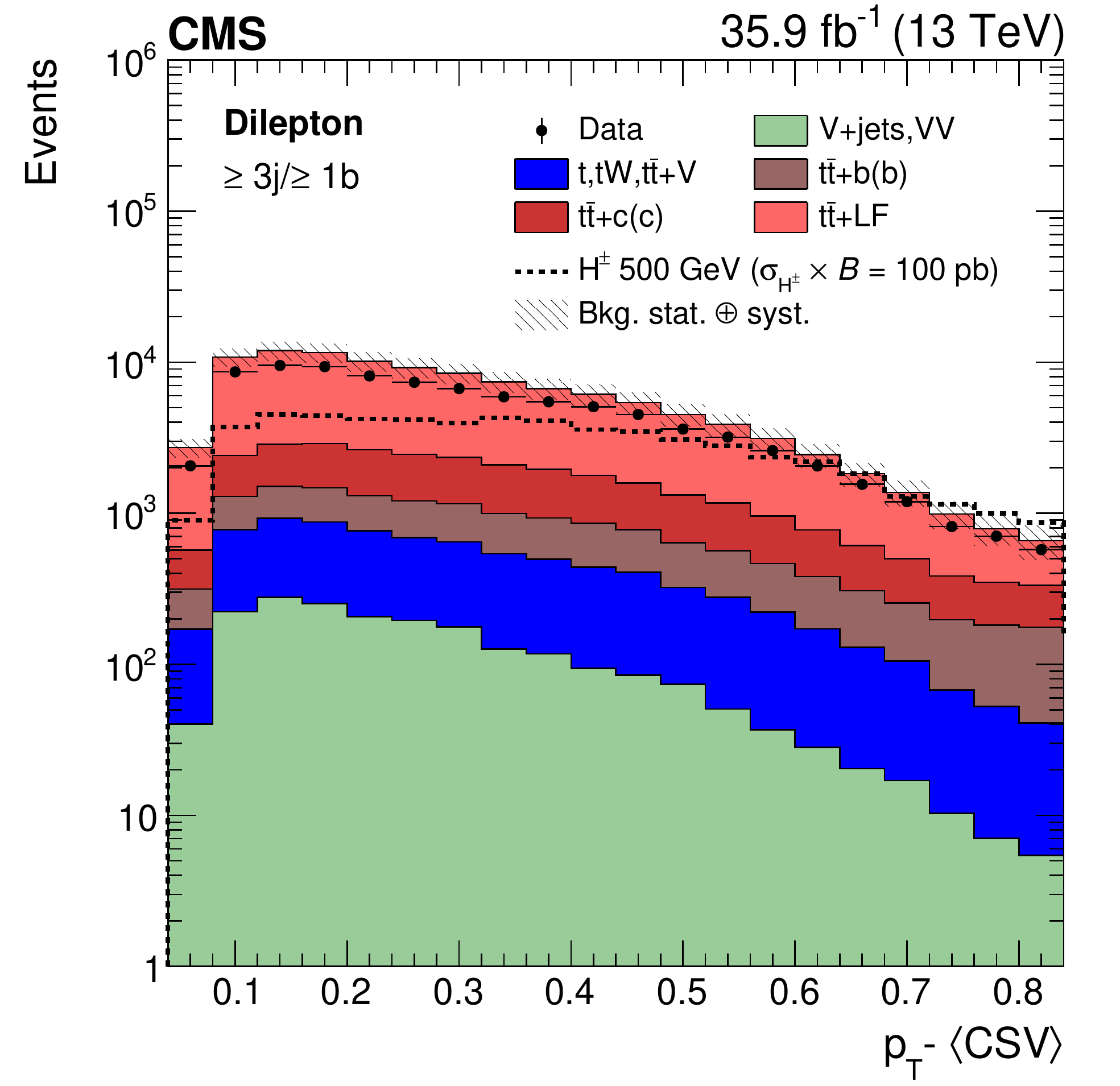}\\
\centering\includegraphics[width=\cmsFigWidth]{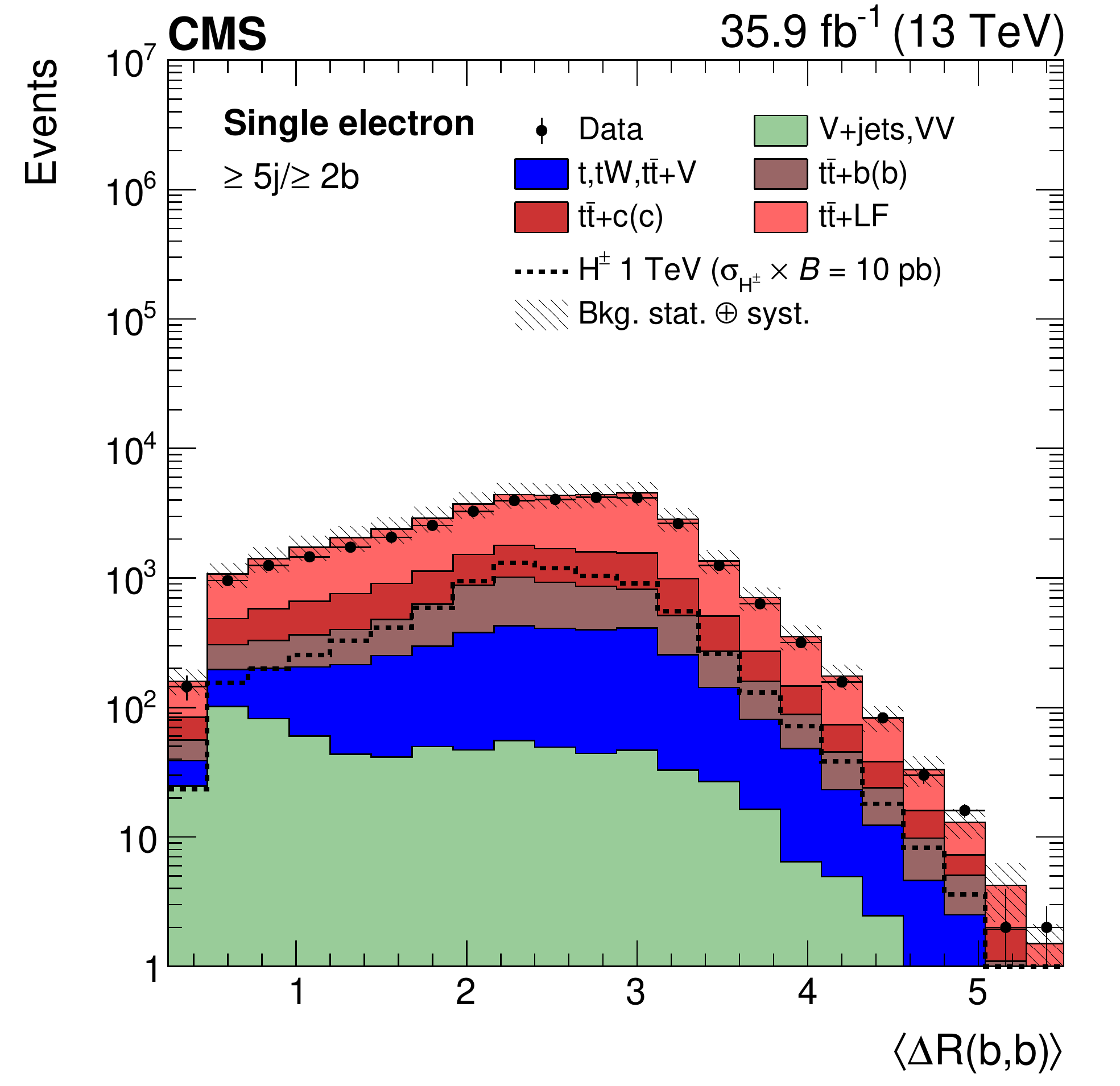}
\centering\includegraphics[width=\cmsFigWidth]{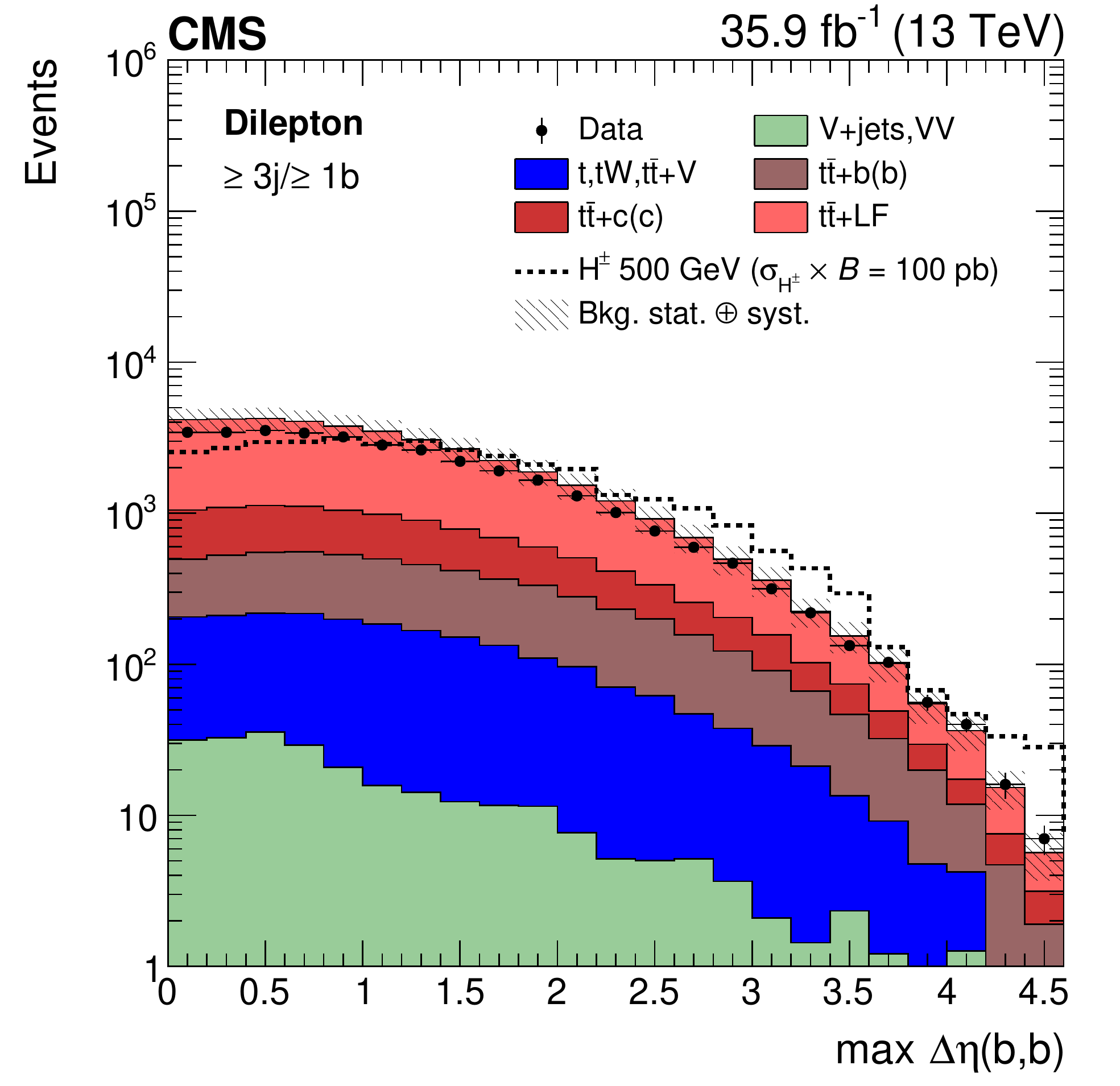}
\caption{
        Representative input variables for the single-lepton $\ge$5j/$\ge$2\cPqb category (\cmsLeft) and for the dilepton $\ge$3j/$\ge$1\cPqb category (\cmsRight) before the signal extraction fit.
        For the single-lepton final state, the centrality (\cmsTop) and $\langle\Delta R(\cPqb,\cPqb)\rangle$ (\cmsBottom) are shown; for the dilepton final state, the $\text{\pt-}\langle\text{CSV}\rangle$ (\cmsTop) and $\max{\Delta\eta(\cPqb,\cPqb)}$ (\cmsBottom) are shown.
    The black markers show the data. The solid histograms represent the background prediction for \ttlf{} (light red), \ttcc{} (dark red), \ttbb{} (brown), single top quark and \ttbar{} in association with extra bosons (blue), and V+jets and multiboson production  (light green).
The dashed line represents the yields for a charged Higgs boson with a mass of 1\TeV(500\GeV) for the single lepton (dilepton) final state and a product of the cross section and branching fraction
of 10 (100)\unit{pb} for the single-lepton and dilepton final states, respectively.
The hatched uncertainty bands include the total uncertainty before the signal extraction fit.
}
\label{fig:controlplots}
\end{figure}

Separate classifiers are constructed for the single-lepton and dilepton final states, using different technologies in order to fully exploit the different sets of features described above.
For each of the suitable discriminating variables, it has been verified that the simulation models data correctly. Figure~\ref{fig:controlplots} shows some of the most important input variables in exemplary signal-region subcategories for the single-lepton ($\ge$5j/$\ge$2\cPqb) and dilepton final states ($\ge$3j/$\ge$1\cPqb).

For all the classifiers described below each signal and background sample is randomly divided into three equally populated parts; one third is used for training the classifiers, one third is used for testing the performance of the classifiers, and one third is used for evaluating the classifier in the context of the maximum-likelihood fit detailed in Section~\ref{sec:results}. The backgrounds are dominated by \ttbar events, but all other SM contributions are also included in the training. Both in the single-lepton and the dilepton regions, the training process and possible sources of over- or under-training are verified by means of statistical tests.

A boosted decision tree (BDT)~\cite{Hastie-Tibshirani-Friedman,Breiman-Friedman-Olshen-Stone} classifier is trained using the TMVA package~\cite{Hocker:2007ht} to discriminate between signal and background in the single-lepton regions.
The dependence of the kinematic signature on \mHpm is accounted for by having a separate training for each \mHpm hypothesis.
The training process is optimized by targeting a region enriched in signal events by requiring $\Njets \geq 5$ and $\Nbjets \geq 2$ (training region).
The binned output distribution of the BDT classifier is calculated in all the single-lepton subcategories corresponding to the training region plus the (4j/$\ge$3\cPqb) region and used in the maximum likelihood fit. In the other single-lepton subcategories, the inclusive event yields are used in the fit to infer additional information on the background normalization.

The dilepton final states exploit a novel technology based on deep neural network (DNN) classifiers~\cite{Hastie-Tibshirani-Friedman},
parametrized as a function of \mHpm~\cite{Baldi:2016fzo}.
The \textsc{TensorFlow} (v1.4.0) backend~\cite{tensorflow} and the \textsc{Keras} (v2.1.1) frontend~\cite{keras} are used to train the classifier.
The parametrization of the signal events as a function of \mHpm enables a unique training for each signal mass hypothesis.
The training process is optimized in the region enriched in signal events by requiring $\Njets \geq 3$ and $\Nbjets \geq 1$.
The jet and \cPqb-tagged jet multiplicities are used
in extending the training parametrization to capture the characteristics of the signal and background processes
in the different regions.
In the regions characterized by a single \cPqb~jet we use the non-tagged jet with the highest value of the \cPqb~tagging discriminator as the second \cPqb~jet for the purpose of computing the input variables.
The binned DNN output is used in the maximum likelihood fit in all the dilepton subcategories to further enhance the separation between the different background processes.

The bin size for the MVA output in each of the subcategories of the analysis is chosen with a variable binning strategy such that the statistical uncertainty in signal and background event yields separately is less than  20\% in each bin.
In order to avoid possible biases in the binning strategy induced by the statistical fluctuations in the simulated samples,
the bin boundaries are defined
based on the events
used for the MVA training.

\section{Background estimation and systematic uncertainties}
\label{sec:bkgSyst}

The leptonic decay of one or two of the \PW~bosons in the \ttbar{} process represents the main background of the analysis for both the single-lepton and dilepton final states.
The \ttbar{} production, as discussed in Section~\ref{sec:datasets}, is separated into \ttlf, \ttbb{}, and \ttcc{} processes.
The last two processes are commonly referred to as $\ttbar$+heavy flavor (HF).
The categorization strategy described in Section~\ref{sec:Overview} populates the low \cPqb~jet multiplicity regions with the \ttlf{} processes, while the regions enriched with the signal are characterized by a larger contribution from the $\cPqt\cPaqt$+HF processes.
Smaller background contributions arise from single top quark production, vector boson production in association with jets, multiboson production processes, \ttbar{} production in association with electroweak bosons (\PW, \PZ, \Pgg, \PH),
and $\cPqt\cPaqt\cPqt\cPaqt$ production.

Different sources of experimental and theoretical uncertainties are modelled as nuisance parameters in the fit and they are allowed to change the event yield,
the migration of events among regions, and the distribution of the MVA output in each category~\cite{COMB-NOTE}.
Uncertainties that purely affect the yield within a category (rate uncertainties)
are modelled via a nuisance parameter with a log-normal probability density function,
while changes in shapes (shape uncertainties) are performed using a polynomial interpolation with a Gaussian constraint,
and they can also change the event yields.
All the sources of systematic uncertainty applied to the analysis are discussed below.

The uncertainty in the integrated luminosity measurement of the 2016 dataset amounts to 2.5\%~\cite{LUM-17-001}.
The uncertainty in the evaluation of the pileup in simulation
is accounted for by varying the total inelastic $\Pp\Pp$ cross section by $\pm$5\%
and propagating the effect of the variation to the final yields.
The difference between the nominal and the altered distributions is taken as the uncertainty and treated as a shape variation in the fit.
Both the integrated luminosity and the pileup uncertainties are separately treated as fully correlated among all processes.

Each reconstructed jet is corrected via calibration factors in order to account for the response of the detectors, with dependencies on the geometry, the pileup conditions, and the kinematic properties of the jet~\cite{CMS-PAS-JME-16-003}.
The uncertainties in the jet energy scale and resolution are propagated by varying the jet momenta and, consequently, the missing transverse momentum.
The events are reanalyzed in order to extract the appropriate rate and shape variations for the final distributions.
An additional uncertainty accounts for the effect of the unclustered energy on \ptmiss.
Each of these uncertainties is treated as fully correlated among all processes.

The \cPqb~tagging and mistagging uncertainties are obtained by varying the corresponding per-jet correction factors within their uncertainties~\cite{Sirunyan:2017ezt}.
The mistag efficiency uncertainties for jets originating from light partons (\cPqu, \cPqd, \cPqs, and \cPg) are considered to be uncorrelated with the \cPqb~tagging efficiency uncertainties, while the $\cPqc$ quark jet mistag rate uncertainties are varied simultaneously with the \cPqb~tagging efficiencies.
The \cPqb~tagging and mistagging efficiency uncertainties are conservatively doubled whenever they are extrapolated outside the $\pt$/$\eta$ range over which the correction factors were derived.
Different sources of uncertainties are varied as independent nuisance parameters.
The portion of the \cPqb~tagging efficiency uncertainty that is correlated with the jet energy scale is evaluated within the overall jet energy scale uncertainty by shifting the \cPqb~tagging scale factors in the same direction as the jet energy scale shift; the procedure reflects the correlation in the derivation of the correction factors.

The uncertainties in the lepton selection efficiency correction factors due to trigger, identification, and isolation efficiencies are applied depending on the lepton \pt{} and $\eta$.
The propagation of the correction factors on the shape of the MVA output impacts only the overall normalization.
The squared sum of the variations due to the identification, isolation, and trigger efficiencies is therefore included as a single rate uncertainty amounting to 3 (4)\% for electrons (muons),
treated as correlated among all the final regions.

Small discrepancies between data and simulation are observed in control regions enriched in processes involving a vector boson with additional jets.
The $\PZ/\Pgg^{*}$ and $\PW$+jets \HT{} distributions are matched to data using corrections derived in a region close to the mass of the \PZ~boson and in the zero \cPqb~jet control region, respectively.
The uncertainties in the derivation of correction factors for the $\PZ/\Pgg^{\scriptscriptstyle *}$ and $\PW$+jets processes in the \HT{} distribution are accounted for in the final results.
They are assumed to be uncorrelated between the two processes and correlated among the analysis regions.

The QCD multijet production is a minor background to the analysis, amounting to about 1\% of the total background across all the signal regions, and is therefore ignored in the fit after the verification of the simulated prediction. For the single-lepton regions, the simulation has been checked in an orthogonal set of events requiring that the \ptmiss is aligned with the jets, while for the dilepton regions, the QCD multijet production  is verified in the same-sign dilepton control regions for each category defined by $\Njets$ and $\Nbjets$.

Theoretical uncertainties related to the PDFs are applied as rate uncertainties to the simulated background samples and account for both the acceptance and the cross section mismodelling~\cite{Butterworth:2015oua}.
Uncertainties from factorization and renormalization scales in the inclusive cross sections are considered independently for each process for which they are non negligible. They are estimated by varying each scale independently from the others by factors of 0.5 and 2 with respect to the default values.

For the simulated samples involving a top quark, an additional uncertainty in the cross sections due to the choice of the top quark mass
is considered by varying the top quark mass by $\pm1.0\GeV$ around the nominal value of 172.5\GeV.

The matching of the \POWHEG\ NLO \ttbar{} matrix element calculation with the \PYTHIA\ parton shower (PS) is varied by shifting the parameter $h_\text{damp}=1.58^{+0.66}_{-0.59} m_{\cPqt}$~\cite{CMS-PAS-TOP-16-021} within the uncertainties. The damping factor $h_\text{damp}$ is used to limit the resummation of higher-order effects by the Sudakov form factor to below a given \pt\,scale~\cite{CMS-PAS-TOP-16-021}.

An additional source of uncertainty arises from the modeling of additional jets by the event generator in top quark pair production.
This uncertainty is estimated in each bin of jet and \cPqb~jet multiplicity, based on the simulated \ttbar{} samples which are enriched or depleted in initial- and final-state radiation. The initial-state radiation PS scale is multiplied by factors of 2 and 0.5 in dedicated simulated samples, whereas the final-state radiation PS scale is scaled up by $\sqrt{2}$ and down by 1/$\sqrt{2}$~\cite{Skands:2014pea,CMS-PAS-TOP-16-021}.
For each PS scale and $h_\text{damp}$ perturbation, the uncertainty is evaluated
as the relative deviation with respect to the nominal event rates.
A nuisance parameter is added for each category defined by $\Njets$ and $\Nbjets$
and considered uncorrelated among regions with different $\Njets$ and also uncorrelated between the single-lepton and dilepton final states.

The normalization of the $\cPqt\cPaqt$+HF processes,
as determined by theoretical calculations~\cite{Jezo:2018yaf} and experimental measurements, is affected by an uncertainty of 50\% that is applied as a rate uncertainty, in addition to the other \ttbar\,cross section uncertainties described above. This procedure allows the signal-depleted regions to determine the overall normalization factor, which includes the production cross section, detector acceptance, and reconstruction efficiencies.

The limited size of the background and signal simulated samples results in statistical fluctuations of the nominal yield prediction.
The content of each bin of each final discriminant distribution is varied by its statistical uncertainty.
The \textit{Barlow--Beeston lite} approach~\cite{BARLOW1993219,Conway:2011in} is applied by assigning, for each bin,
the combined statistical uncertainty of all simulated samples to the process dominating the background yield in that bin.
Since all bins are statistically independent, each variation is treated as uncorrelated with any other variation.

A summary of the effects of the systematic uncertainties on the event yields, summed over all final states and regions, is provided prior to the fit to data in Table~\ref{tab:syst}.

\begin{table}[ht]
    \centering
    \topcaption{Effects of the systematic uncertainties as the variation (in percent) of the event yields prior to the fit to data, summed over all final states and regions. The column \textit{Shape} reports whether a given uncertainty is considered a shape uncertainty or a rate uncertainty.
    }
    \label{tab:syst}
\cmsTable{
\begin{tabular}{ l c *{6}{x{-1}}}
    \mc{Source of uncertainty} & \mc{Shape}  & \mc{$\PH^\pm$} & \mc{\ttlf} & \mc{\ttcc} & \mc{\ttbb} & \mc{$\cPqt$, $\cPqt\PW$, $\ttbar$+X} & \mc{V+jets}\\
    \hline
    Integrated luminosity & & 2.5 & 2.5 & 2.5 & 2.5 & 2.5 & 2.5 \\
    Pileup & $\checkmark$ & 0.4 & 0.2 & 0.2 & 0.2 & 0.2 & 1.8 \\
    Jet energy scale and resolution & $\checkmark$ & 2.8 & 3.9 & 3.3 & 3.0 & 3.9 & 5.3 \\
    \cPqb~jet identification & $\checkmark$ & 4.6 & 3.1 & 4.1 & 4.6 & 3.0 & 11.6 \\
    Lepton selection efficiency & & 3.4 & 3.1 & 3.3 & 3.3 & 3.3 & 3.7 \\
    Unclustered \ptmiss energy &  & 2.0 & 2.0 & 2.0 & 2.0 & 2.0 & 2.0 \\
    Acceptance (scales, PDF) & $\checkmark$ & 9.8 & 9.0 & 11.4 & 12.0 & 3.3 & 11.2 \\
    Cross section (scales, PDF) & & \mc{\NA} & 5.5 & 5.5 & 5.5 & 4.0 & 4.1 \\
    Top quark mass & & \mc{\NA} & 2.7 & 2.7 & 2.7 & 2.2 & \mc{\NA} \\
    \ttbar parton showering & & \mc{\NA} & 6.4 & 10.6 & 9.5 & \mc{\NA} & \mc{\NA} \\
    \ttbar{}+HF normalization & & \mc{\NA} & \mc{\NA} & 50.0 & 50.0 & \mc{\NA} & \mc{\NA} \\
\end{tabular}
}
\end{table}

\section{Results}
\label{sec:results}

The statistical interpretation is based on a simultaneous fit of the MVA output discriminators
and event yields in the different signal regions described in Section~\ref{sec:Overview}.
The parameter of interest reflecting the signal normalization $\sPPtoHtb\BHtb=\sPPtoHpBoth\BHptb+\sPPtoHmBoth\BHmtb$ and the nuisance parameters specified in Section~\ref{sec:bkgSyst}
are encoded in the negative log-likelihood function and profiled in the minimization process.
The log-likelihood ratio is used as test statistic to assess
the agreement of data with the background-only hypothesis or the presence of the signal
and the asymptotic approximation is used in the statistical analysis~\cite{Cowan:2010js,COMB-NOTE}.
The statistical method used to report the results is the \CLs{} modified frequentist criterion~\cite{CLS1,CLS2}.

\begin{figure}[h!p]
\centering\includegraphics[width=\cmsFigWidth]{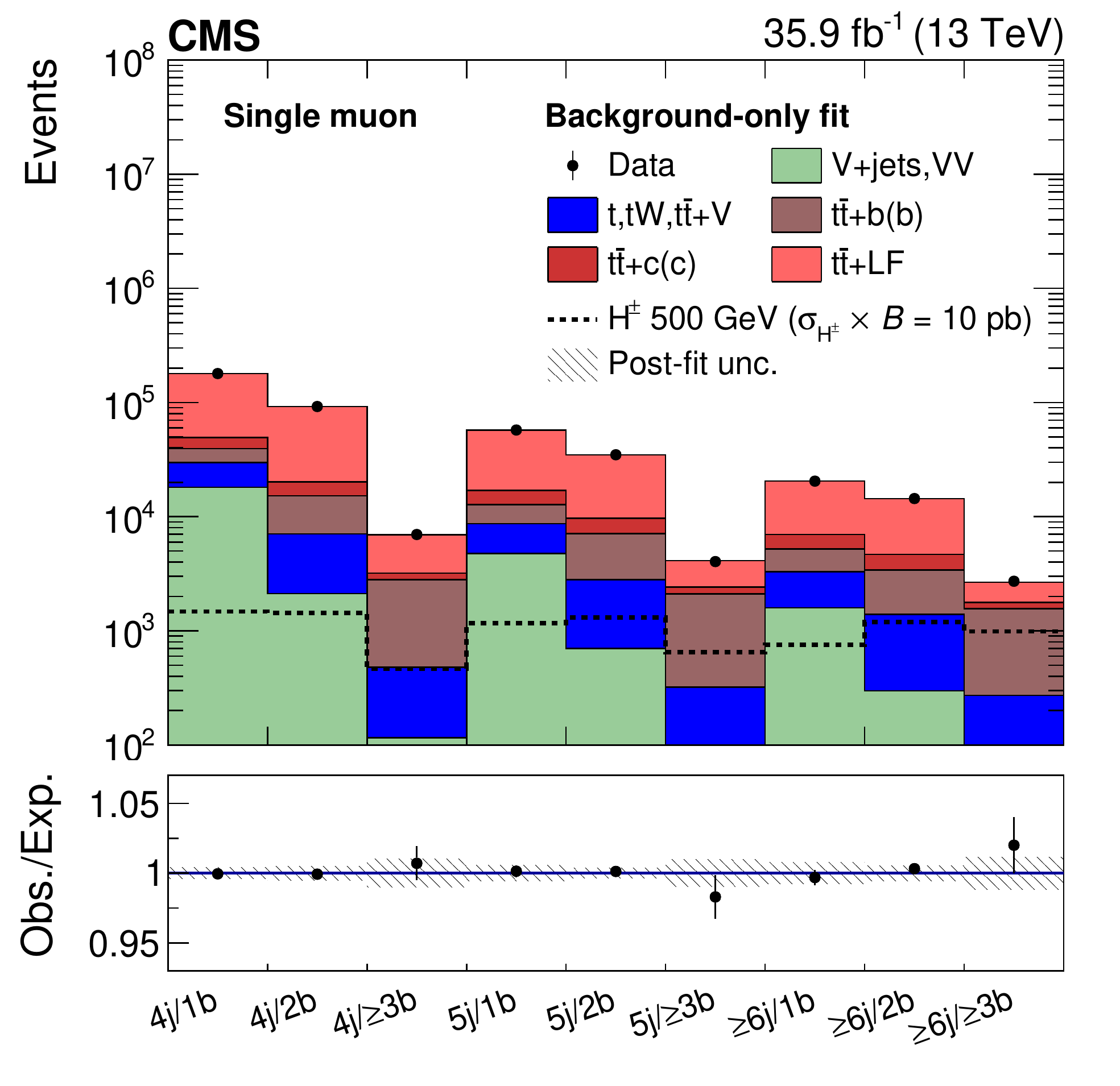}
\centering\includegraphics[width=\cmsFigWidth]{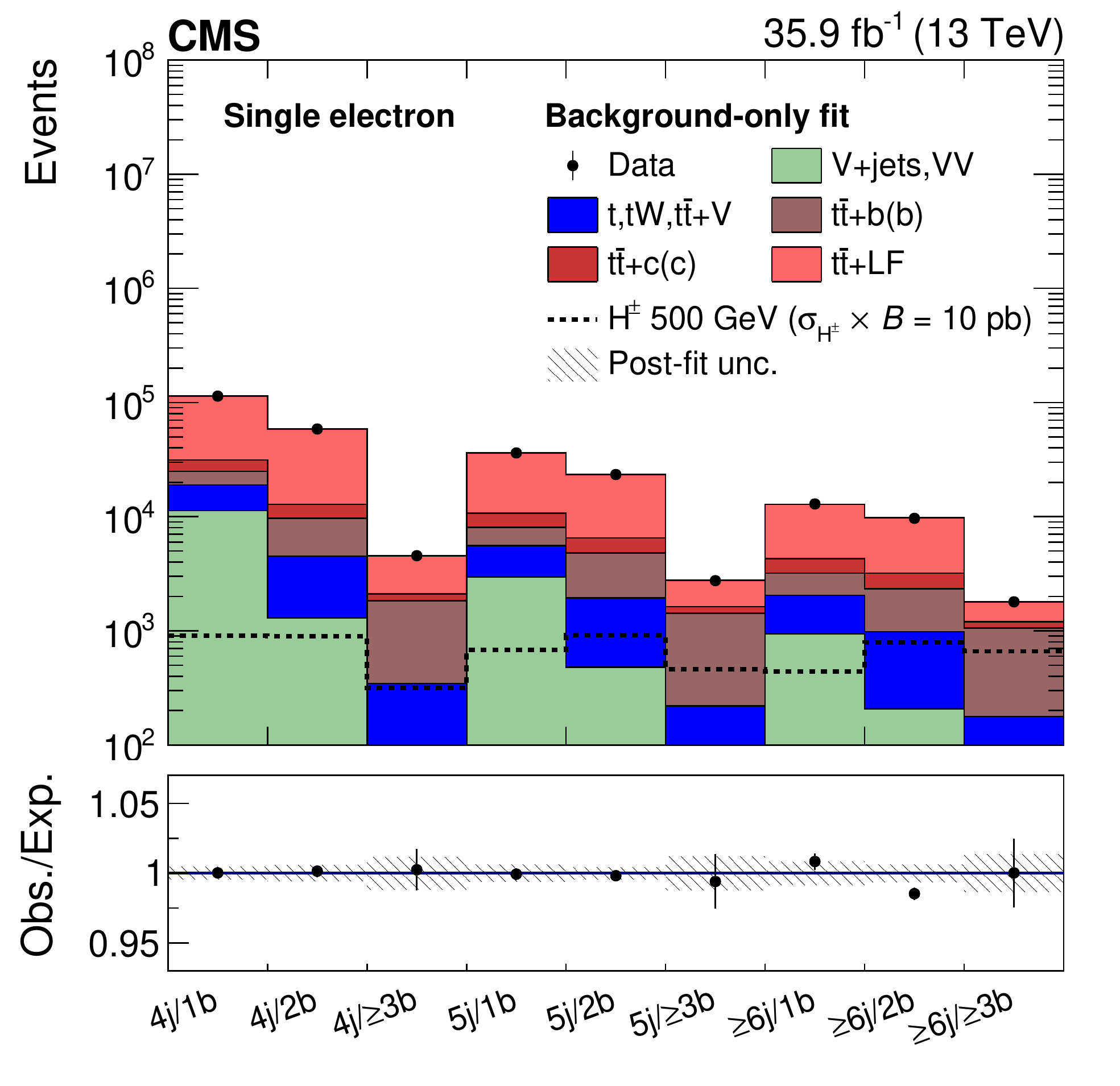}
\centering\includegraphics[width=\cmsFigWidth]{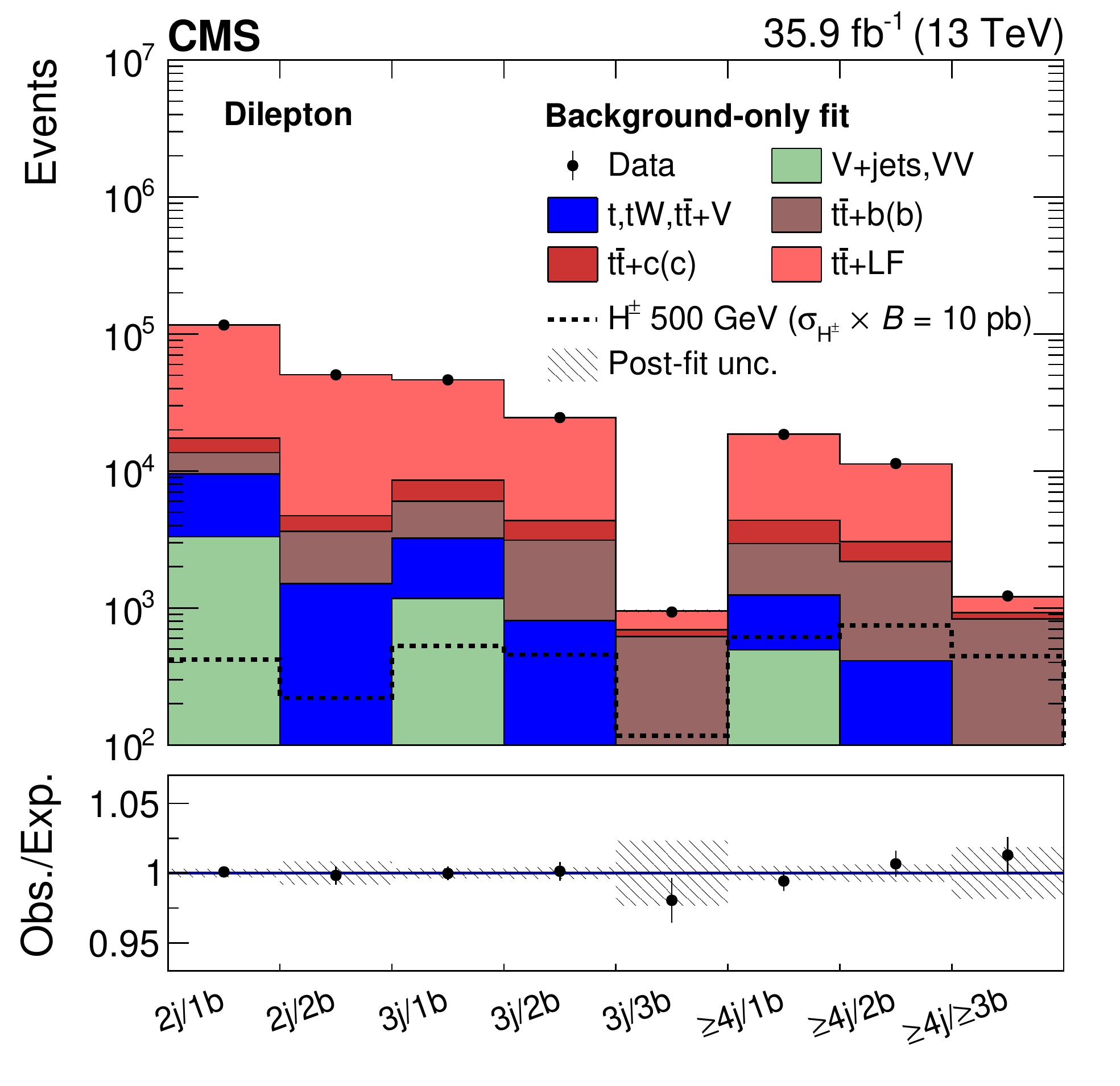}
\caption{
    Summary of event yields in each analysis category for single-muon (\cmsTop \cmsLeft), single-electron (\cmsTop \cmsRight), and dilepton (\cmsBottom) final states.
The yields observed in data (black markers) are overlaid. The solid histograms represent the background prediction for \ttlf{} (light red), \ttcc{} (dark red), \ttbb{} (brown), single top quark and \ttbar{} in association with extra bosons (blue), and V+jets and multiboson production  (light green).
The dashed line represents the yields for a charged Higgs boson with a mass of 500\GeV and
a product of the cross section and the branching fraction
of 10\unit{pb}.
The lower panel shows the ratio of data to the SM expectation after the background-only fit to the data and the hatched uncertainty bands include the total uncertainty.
}
\label{fig:summary}
\end{figure}

\begin{figure}[h!p]
\centering\includegraphics[width=\cmsFigWidth]{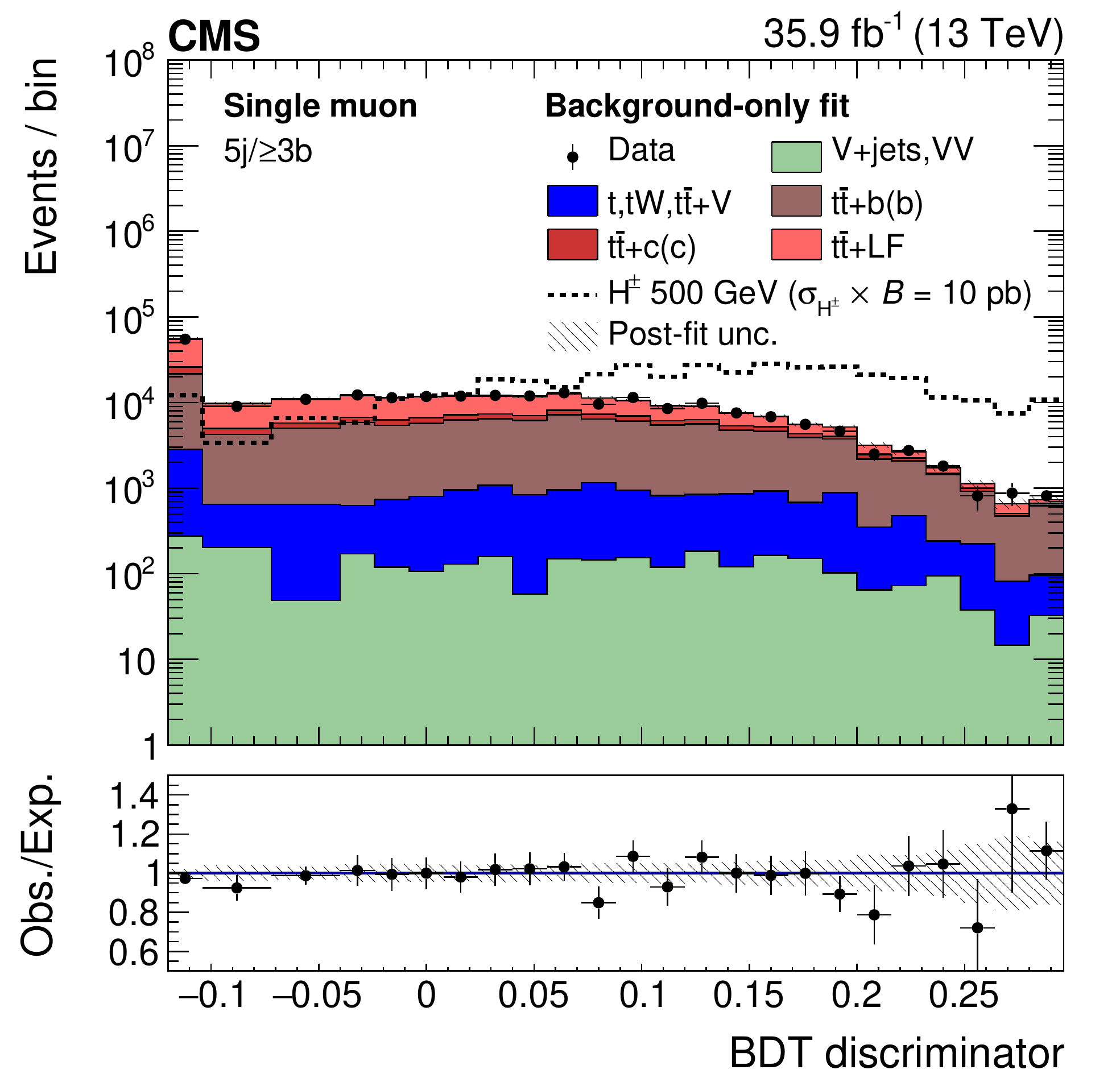}
\centering\includegraphics[width=\cmsFigWidth]{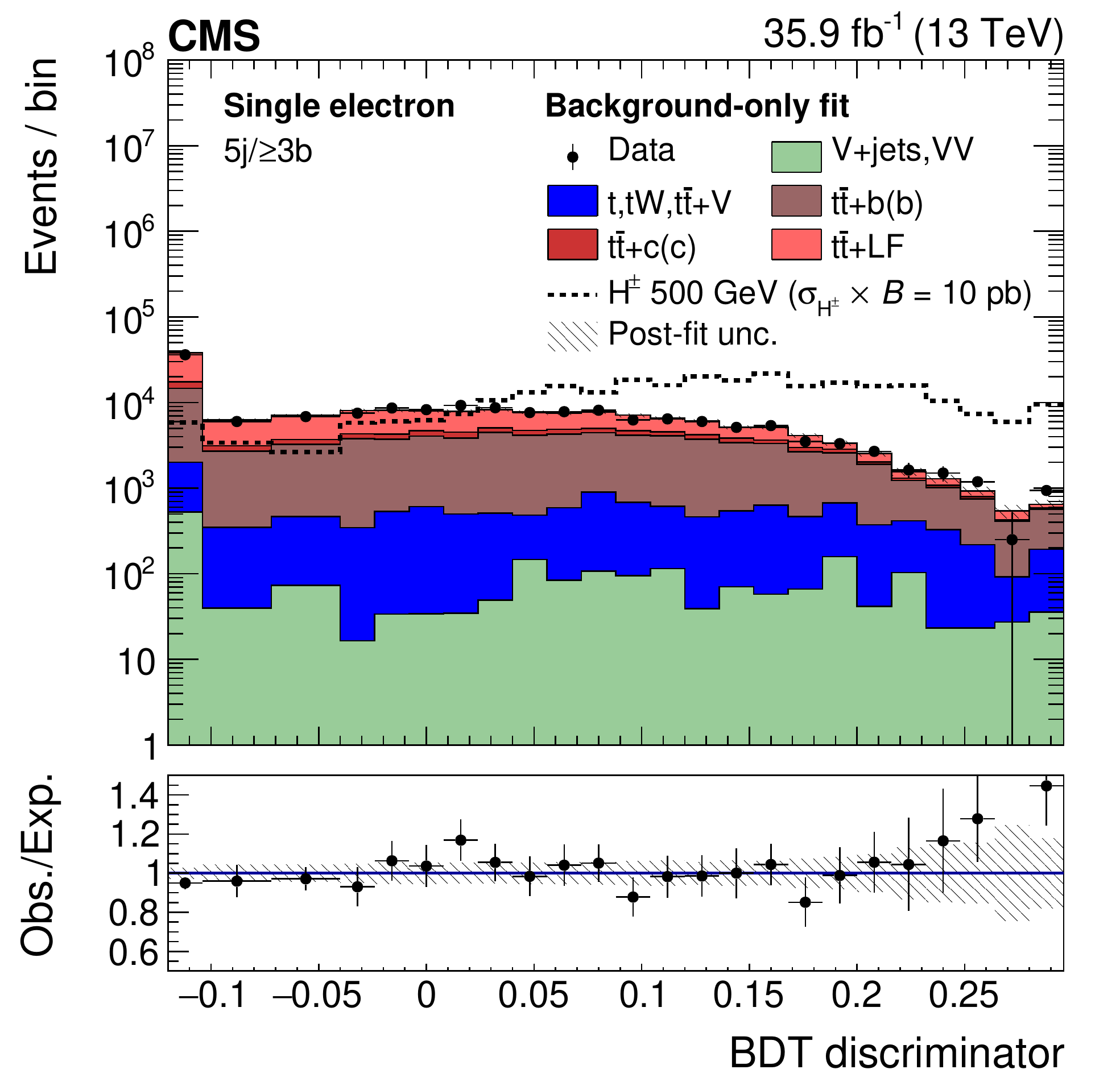}
\centering\includegraphics[width=\cmsFigWidth]{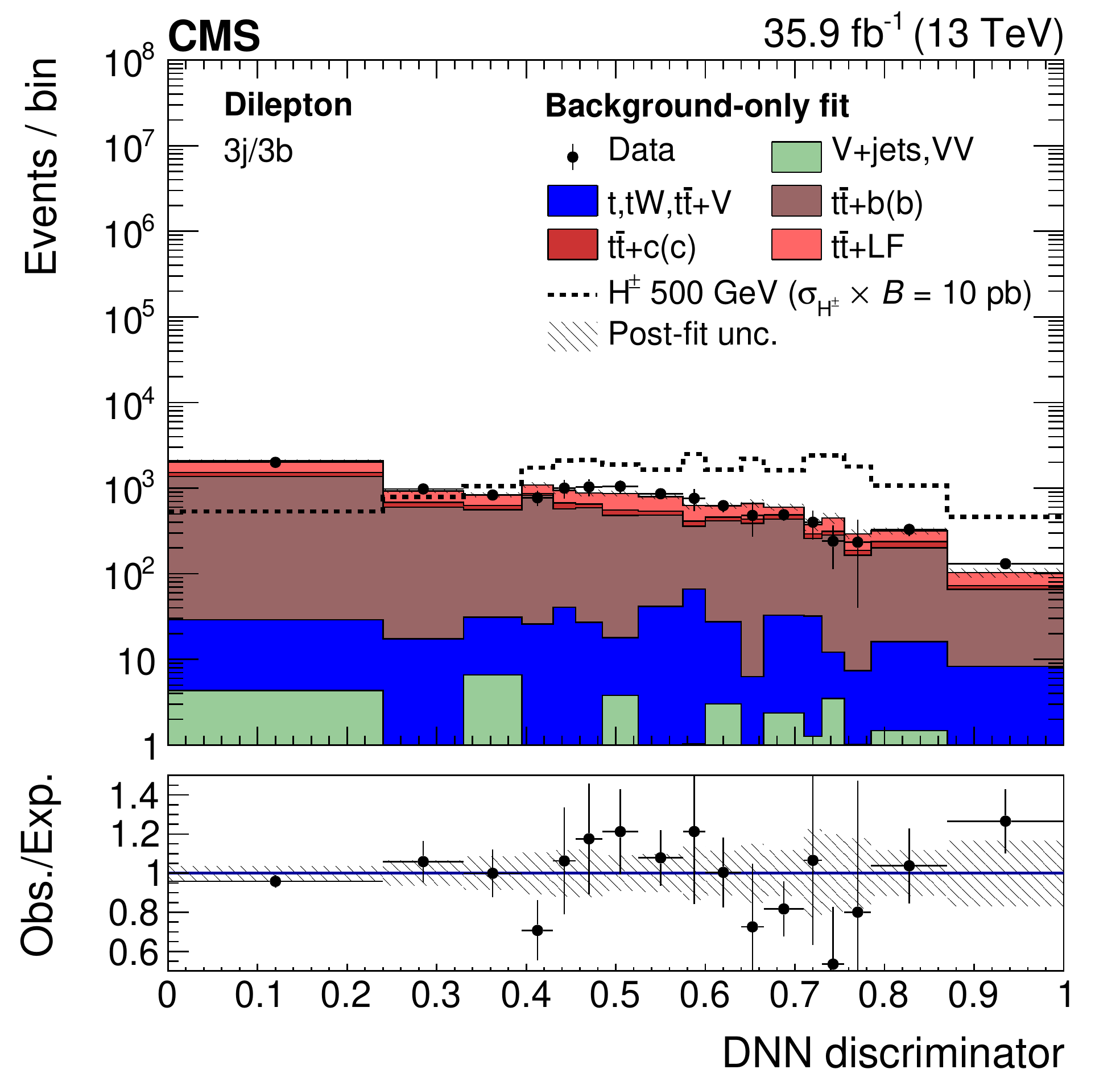}
\caption{
    Distributions of the MVA outputs of the data and the SM expectation after the background-only fit to the data for the single-muon 5j/$\ge$3\cPqb category (\cmsTop \cmsLeft), for the single-electron 5j/$\ge$3\cPqb category (\cmsTop \cmsRight), and for the dilepton 3j/3\cPqb category (\cmsBottom).
    The black markers show the data. The solid histograms represent the background prediction for \ttlf{} (light red), \ttcc{} (dark red), \ttbb{} (brown), single top quark and \ttbar{} in association with extra bosons (blue), and V+jets and multiboson production  (light green).
The dashed line represents the yields for a charged Higgs boson with a mass of 500\GeV and
a product of the cross section and branching fraction
of 10\unit{pb}.
The lower panel shows the ratio of data to the SM expectation after the background-only fit to the data and the hatched uncertainty bands include the total uncertainty.
}
\label{fig:MVAtemplatesSR}
\end{figure}

Figure~\ref{fig:summary} shows the event yields in the subcategories of the analysis after a background-only fit to data.
In the regions where the shape of the MVA classifier output is used, the yields are obtained by integrating the distribution and the correlations across the bins are accounted for in the quoted uncertainties. The contribution of a hypothetical charged Higgs boson with a mass of 500\GeV and $\sPPtoHtb\,\BHtb=10$\unit{pb} is also displayed.
In the same configuration,
Fig.~\ref{fig:MVAtemplatesSR} shows the MVA (BDT and DNN) outputs in exemplary signal-region subcategories for the single-lepton (5j/$\ge$3\cPqb) and dilepton (3j/3\cPqb) final states.

\begin{figure}[htbp]
\centering\includegraphics[width=\cmsFigWidth]{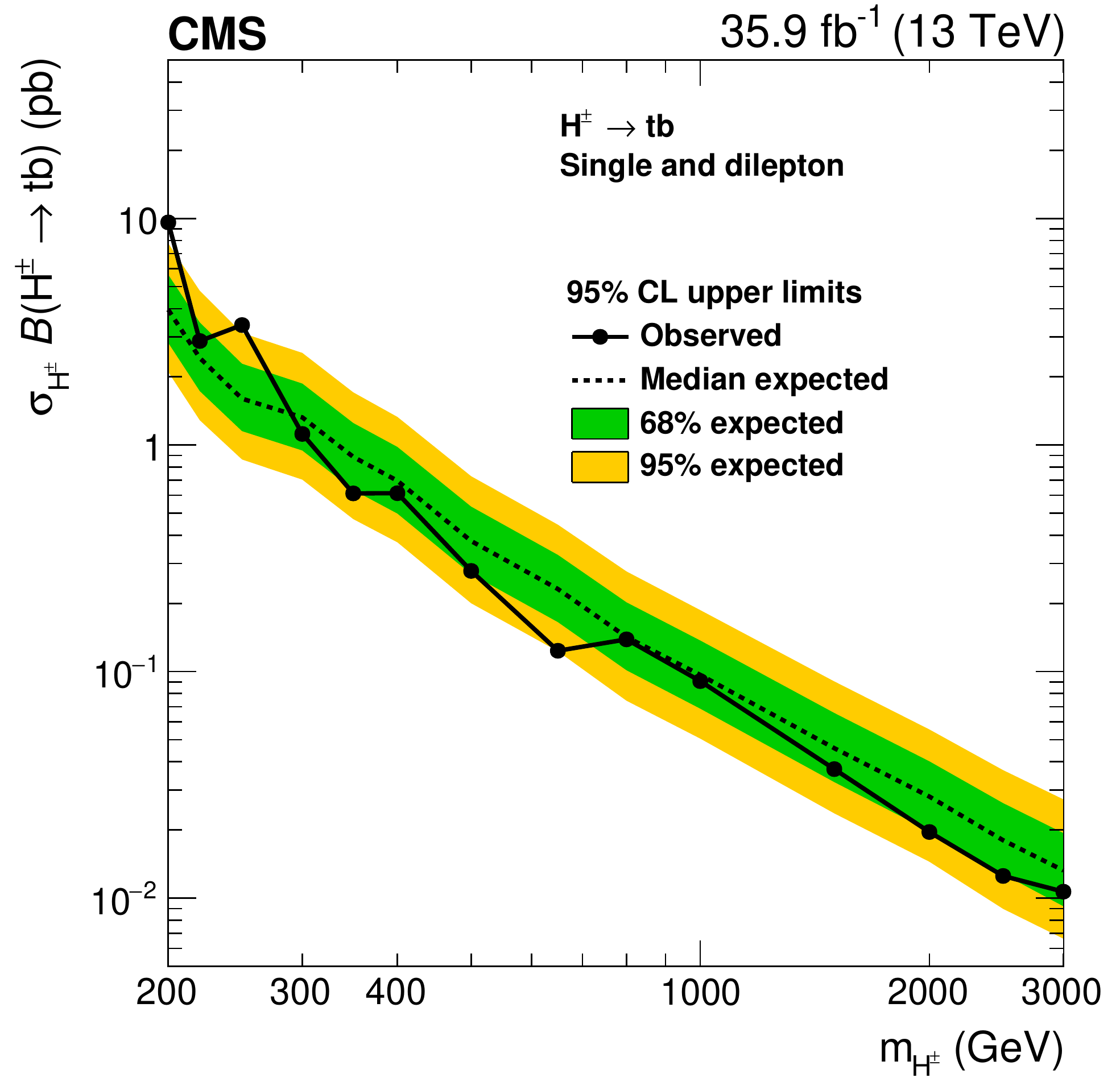}
\centering\includegraphics[width=\cmsFigWidth]{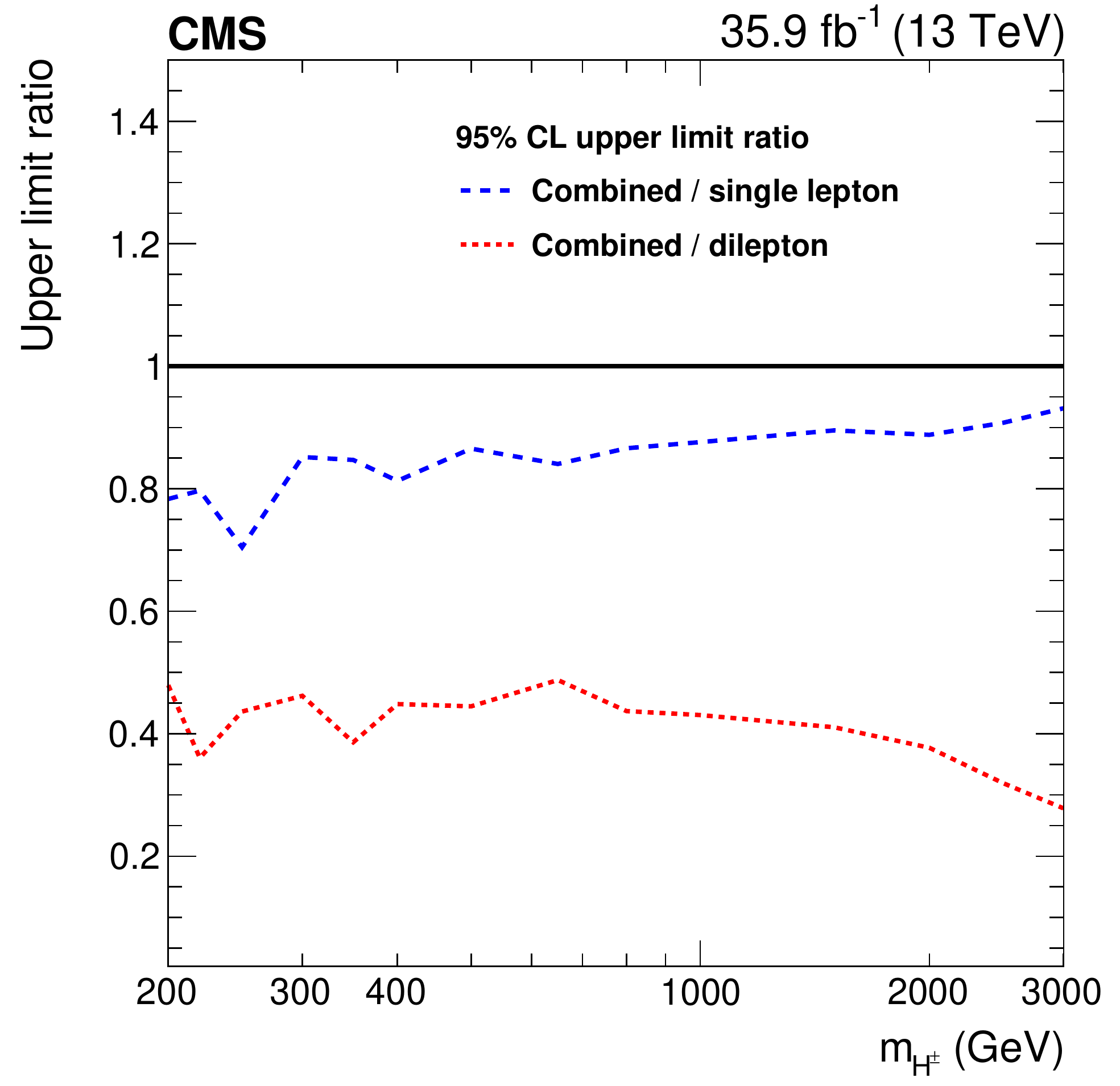}
\caption{
    The upper limit at 95\% \CL on $\sPPtoHtb\BHtb$
    with single-lepton and dilepton final states combined (\cmsLeft).
    The solid black markers describe the observed upper limits, while the dashed line corresponds to the expectations from the background.
    The green (yellow) band represents one (two) standard deviations from the expected median.
    The contribution of the single-lepton and dilepton regions to the combined limit is also represented, expressed as a ratio (\cmsRight).
}
\label{fig:Combination}
\end{figure}

The data agree with the background distributions and no significant excess is observed.
Exclusion limits are set at 95\% confidence level (\CL) on $\sPPtoHtb\BHtb$
for \mHpm hypotheses between 200 and 3000\GeV.
The observed (expected) upper limits with single-lepton and dilepton final states combined are shown in Fig.~\ref{fig:Combination} (\cmsLeft) and  listed in Table~\ref{tab:Limit_12L}.
The single-lepton and dilepton regions have comparable sensitivity in the low-mass regime ($\approx$200\GeV) while the single-lepton regions become increasingly dominant at higher values of the mass hypothesis; Figure~\ref{fig:Combination} (\cmsRight) details the contributions of the single-lepton and dilepton regions.
Using the MVA classifier instead of the \HT{}-based approach of the previous publication~\cite{Khachatryan:2015qxa} yields an improvement of 20--40\% in the expected limits, depending on the signal mass.

\begin{table}[h]
\centering
\topcaption{The upper limits at 95\% \CL on the $\sPPtoHtb\BHtb$ with the single-lepton and dilepton final states combined. The one (two) standard deviations (s.d.) from the expected median are also reported.
}
\cmsTable{
\begin{tabular}{ *{7}{x{-1}} }
\mc{\multirow{2}{*}{\mHpm (\GeVns{})}}  & \multicolumn{5}{c}{Expected limits (pb)} & \mc{\multirow{2}{*}{Observed limits (pb)}} \\
\cline{2-6}
 & \mc{-2 s.d.} & \mc{-1 s.d.} & \mc{median expected}  &\mc{+1 s.d.}  & \mc{+2 s.d.} & \\
\hline
200 & 2.1 & 2.8 & 4.0 & 5.6 & 7.8 & 9.6 \\
220 & 1.3 & 1.7 & 2.4 & 3.5 & 4.8 & 2.9 \\
250 & 0.9 & 1.2 & 1.6 & 2.3 & 3.1 & 3.4 \\
300 & 0.7 & 0.9 & 1.3 & 1.9 & 2.6 & 1.1 \\
350 & 0.47 & 0.63 & 0.89 & 1.25 & 1.71 & 0.61 \\
400 & 0.37 & 0.50 & 0.70 & 0.98 & 1.33 & 0.61 \\
500 & 0.20 & 0.27 & 0.38 & 0.53 & 0.73 & 0.28 \\
650 & 0.12 & 0.17 & 0.23 & 0.33 & 0.45 & 0.12 \\
800 & 0.07 & 0.10 & 0.14 & 0.20 & 0.28 & 0.14 \\
1000 & 0.051 & 0.069 & 0.097 & 0.137 & 0.187 & 0.091 \\
1500 & 0.024 & 0.033 & 0.046 & 0.066 & 0.090 & 0.037 \\
2000 & 0.015 & 0.020 & 0.028 & 0.040 & 0.056 & 0.020 \\
2500 & 0.009 & 0.013 & 0.018 & 0.026 & 0.037 & 0.013 \\
3000 & 0.007 & 0.009 & 0.013 & 0.019 & 0.027 & 0.011 \\
\end{tabular}
}
\label{tab:Limit_12L}
\end{table}

\begin{figure}[htp]
    \centering
    \includegraphics[width=0.49\textwidth]{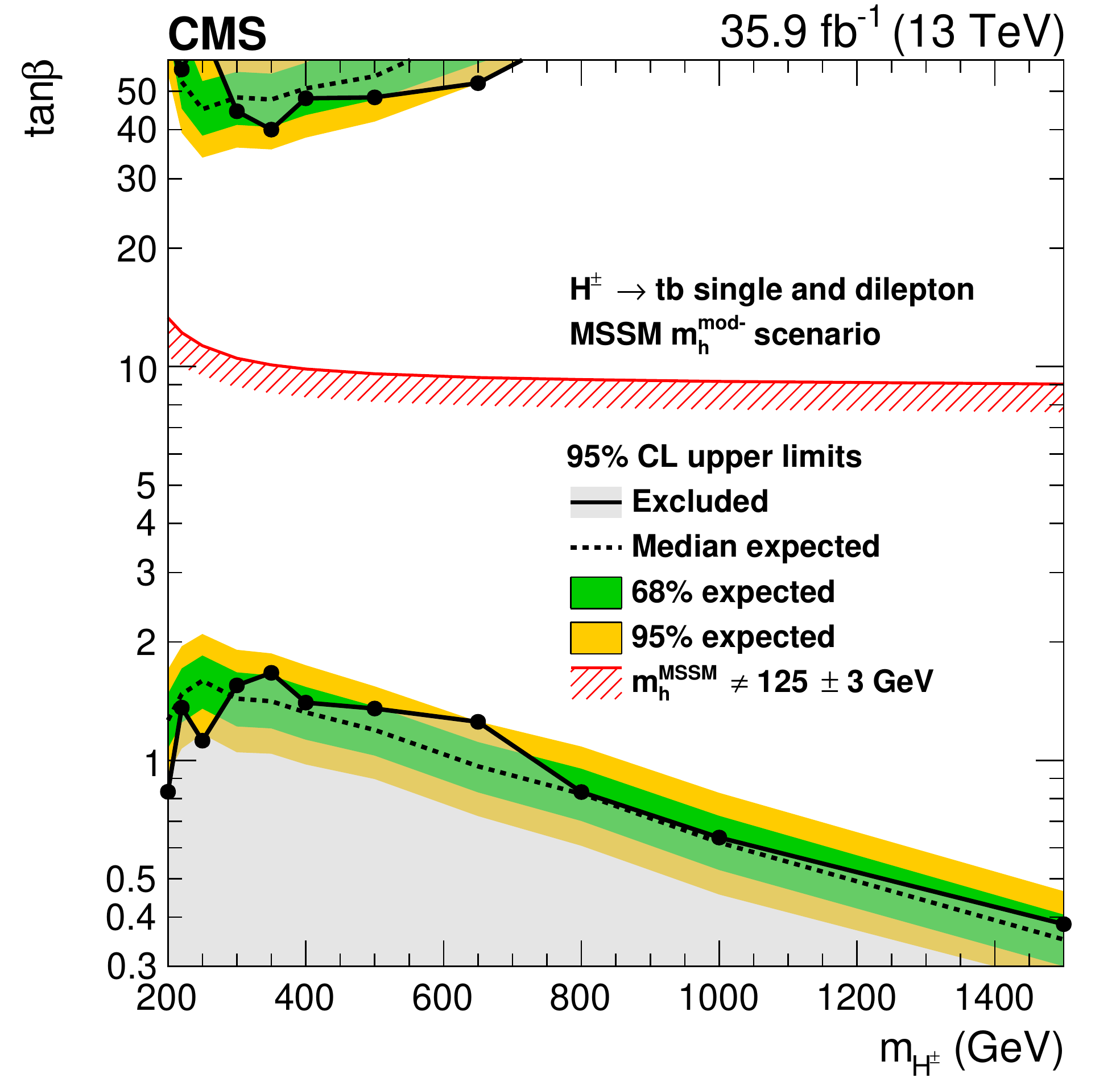}
    \includegraphics[width=0.49\textwidth]{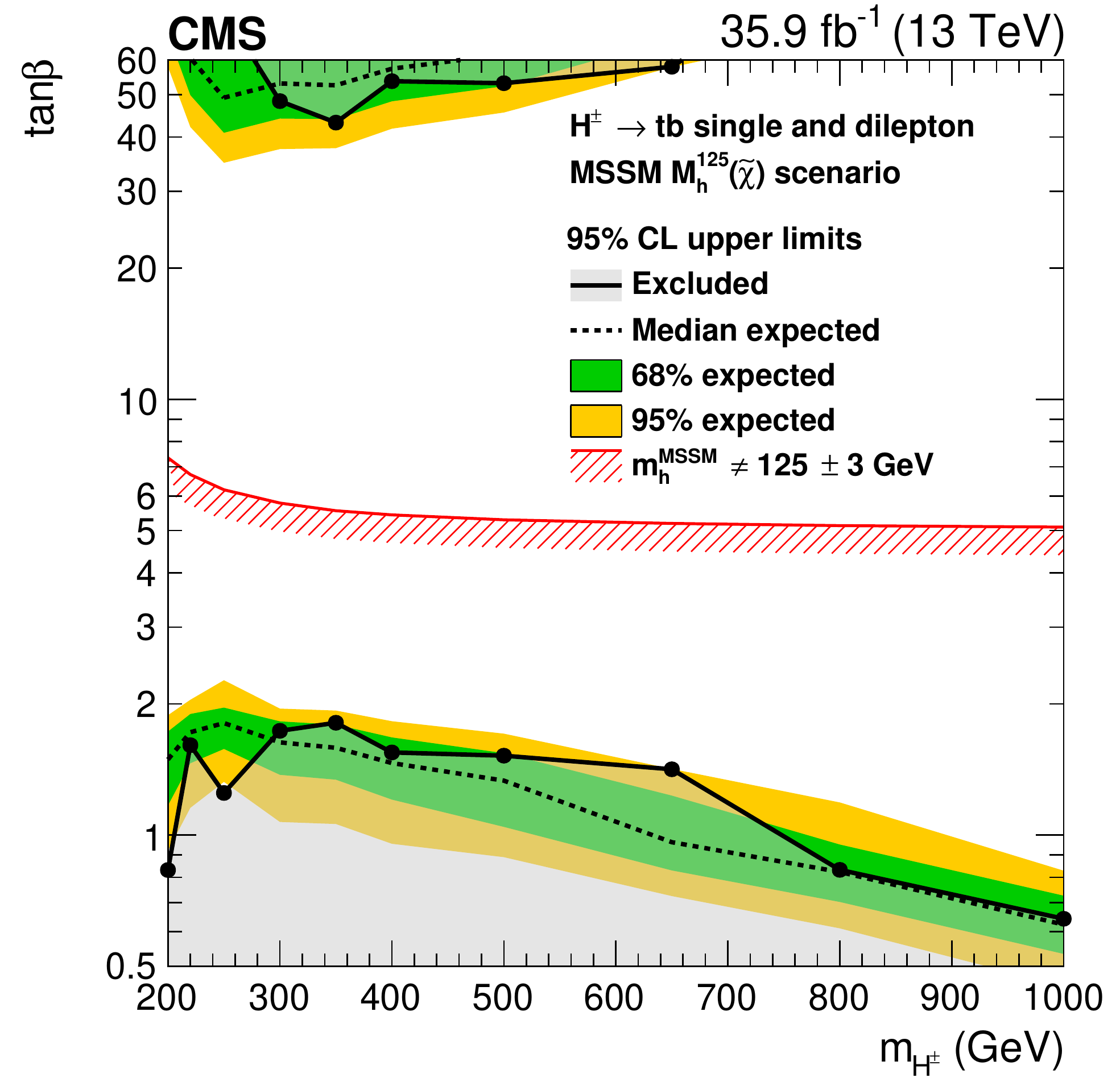}
    \caption{Excluded parameter space regions in the \mhmodm{}scenario (\cmsLeft) and in the \mhonetwentyfive{} scenario (\cmsRight).
        The grey area delimited by the solid black line and markers represents the observed excluded region.
        The dashed black line and the green (yellow) regions represent the median expected exclusion regions and one (two) standard deviations from the expected median, respectively. The region below the red line is excluded assuming that the observed neutral Higgs boson is the light CP-even 2HDM Higgs boson with a mass of ${125\pm3\GeV}$, where the uncertainty is the theoretical uncertainty in the mass calculation.
    }
    \label{fig:moddep}
\end{figure}

The model-dependent upper limits are obtained by comparing the observed limits with the theoretical predictions.
The MSSM \mhmodm{} benchmark scenario~\cite{Carena:2013ytb} is designed to give a mass of approximately 125\GeV for the light CP-even 2HDM Higgs boson over a wide region of the parameter space.
The \mhonetwentyfive{} scenario~\cite{Bahl:2018zmf} is characterized by small gaugino and Higgs/higgsino superpotential masses which are also close to each other; this results in a significant mixing parameter between higgsinos and gauginos and in a compressed electroweakino mass spectrum. The phenomenology of the \mhonetwentyfive{} scenario resembles therefore the Type-II 2HDM with MSSM-inspired Higgs couplings compatible with $\mh\approx125\GeV$ for large masses of the pseudoscalar boson, A.
Figure~\ref{fig:moddep} shows the excluded parameter space in the MSSM \mhmodm{} and \mhonetwentyfive{} scenarios.
In both models, the observed exclusion of high values of \tanbeta is in the range 40--60 in the \mHpm range of 200--700\GeV;
for low values of \tanbeta the values 0.4--1.5 are excluded in the \mHpm range of 200\GeV--1.5\TeV in the context of \mhmodm{} scenario while
the values 0.6--1.5 are excluded in the \mHpm range of 200\GeV--1\TeV for the \mhonetwentyfive{} scenario.

\section{Summary}

A search is presented for a charged Higgs boson decaying into a top-bottom quark-antiquark pair when produced in association with a top quark or a top and a bottom quark.
The analyzed proton-proton collision data, collected at \sqrtsThirteen with the CMS detector at the LHC, correspond to an integrated luminosity of \thelumi.
The search uses events with a single isolated electron or muon or an opposite-sign electron or muon pair.
Events are categorized according to the jet multiplicity and the number of jets identified as containing a \cPqb-hadron decay.
Multivariate techniques are used to discriminate between signal and background events, the latter being dominated by \ttbar{} production.
Results are presented for a charged Higgs boson with a mass larger than the top quark mass.
95\% confidence level upper limits of 9.6--0.01\unit{pb} are set
on the product of the charged Higgs boson production cross section and the branching fraction into a top-bottom quark-antiquark pair, $\sPPtoHtb\BHtb=\sPPtoHpBoth\BHptb+\sPPtoHmBoth\BHmtb$,
in the mass range from 200\GeV to 3\TeV, representing an improvement over previous results~\cite{Khachatryan:2015qxa,Aad:2015typ,Aaboud:2018cwk} by a factor of about 2--7 in the given mass range.
Exclusion regions in the parameter space of the minimal supersymmetric standard model \mhmodm and \mhonetwentyfive{} benchmark scenarios are presented.

\begin{acknowledgments}
We congratulate our colleagues in the CERN accelerator departments for the excellent performance of the LHC and thank the technical and administrative staffs at CERN and at other CMS institutes for their contributions to the success of the CMS effort. In addition, we gratefully acknowledge the computing centers and personnel of the Worldwide LHC Computing Grid for delivering so effectively the computing infrastructure essential to our analyses. Finally, we acknowledge the enduring support for the construction and operation of the LHC and the CMS detector provided by the following funding agencies: BMBWF and FWF (Austria); FNRS and FWO (Belgium); CNPq, CAPES, FAPERJ, FAPERGS, and FAPESP (Brazil); MES (Bulgaria); CERN; CAS, MoST, and NSFC (China); COLCIENCIAS (Colombia); MSES and CSF (Croatia); RPF (Cyprus); SENESCYT (Ecuador); MoER, ERC IUT, PUT and ERDF (Estonia); Academy of Finland, MEC, and HIP (Finland); CEA and CNRS/IN2P3 (France); BMBF, DFG, and HGF (Germany); GSRT (Greece); NKFIA (Hungary); DAE and DST (India); IPM (Iran); SFI (Ireland); INFN (Italy); MSIP and NRF (Republic of Korea); MES (Latvia); LAS (Lithuania); MOE and UM (Malaysia); BUAP, CINVESTAV, CONACYT, LNS, SEP, and UASLP-FAI (Mexico); MOS (Montenegro); MBIE (New Zealand); PAEC (Pakistan); MSHE and NSC (Poland); FCT (Portugal); JINR (Dubna); MON, RosAtom, RAS, RFBR, and NRC KI (Russia); MESTD (Serbia); SEIDI, CPAN, PCTI, and FEDER (Spain); MOSTR (Sri Lanka); Swiss Funding Agencies (Switzerland); MST (Taipei); ThEPCenter, IPST, STAR, and NSTDA (Thailand); TUBITAK and TAEK (Turkey); NASU and SFFR (Ukraine); STFC (United Kingdom); DOE and NSF (USA).

\hyphenation{Rachada-pisek} Individuals have received support from the Marie-Curie program and the European Research Council and Horizon 2020 Grant, contract Nos.\ 675440, 752730, and 765710 (European Union); the Leventis Foundation; the A.P.\ Sloan Foundation; the Alexander von Humboldt Foundation; the Belgian Federal Science Policy Office; the Fonds pour la Formation \`a la Recherche dans l'Industrie et dans l'Agriculture (FRIA-Belgium); the Agentschap voor Innovatie door Wetenschap en Technologie (IWT-Belgium); the F.R.S.-FNRS and FWO (Belgium) under the ``Excellence of Science -- EOS" -- be.h project n.\ 30820817; the Beijing Municipal Science \& Technology Commission, No. Z181100004218003; the Ministry of Education, Youth and Sports (MEYS) of the Czech Republic; the Lend\"ulet (``Momentum") Program and the J\'anos Bolyai Research Scholarship of the Hungarian Academy of Sciences, the New National Excellence Program \'UNKP, the NKFIA research grants 123842, 123959, 124845, 124850, 125105, 128713, 128786, and 129058 (Hungary); the Council of Science and Industrial Research, India; the HOMING PLUS program of the Foundation for Polish Science, cofinanced from European Union, Regional Development Fund, the Mobility Plus program of the Ministry of Science and Higher Education, the National Science Center (Poland), contracts Harmonia 2014/14/M/ST2/00428, Opus 2014/13/B/ST2/02543, 2014/15/B/ST2/03998, and 2015/19/B/ST2/02861, Sonata-bis 2012/07/E/ST2/01406; the National Priorities Research Program by Qatar National Research Fund; the Ministry of Science and Education, grant no. 3.2989.2017 (Russia); the Programa Estatal de Fomento de la Investigaci{\'o}n Cient{\'i}fica y T{\'e}cnica de Excelencia Mar\'{\i}a de Maeztu, grant MDM-2015-0509 and the Programa Severo Ochoa del Principado de Asturias; the Thalis and Aristeia programs cofinanced by EU-ESF and the Greek NSRF; the Rachadapisek Sompot Fund for Postdoctoral Fellowship, Chulalongkorn University and the Chulalongkorn Academic into Its 2nd Century Project Advancement Project (Thailand); the Welch Foundation, contract C-1845; and the Weston Havens Foundation (USA).
\end{acknowledgments}

\clearpage

\bibliography{auto_generated}

\cleardoublepage \appendix\section{The CMS Collaboration \label{app:collab}}\begin{sloppypar}\hyphenpenalty=5000\widowpenalty=500\clubpenalty=5000\vskip\cmsinstskip
\textbf{Yerevan Physics Institute, Yerevan, Armenia}\\*[0pt]
A.M.~Sirunyan$^{\textrm{\dag}}$, A.~Tumasyan
\vskip\cmsinstskip
\textbf{Institut f\"{u}r Hochenergiephysik, Wien, Austria}\\*[0pt]
W.~Adam, F.~Ambrogi, T.~Bergauer, J.~Brandstetter, M.~Dragicevic, J.~Er\"{o}, A.~Escalante~Del~Valle, M.~Flechl, R.~Fr\"{u}hwirth\cmsAuthorMark{1}, M.~Jeitler\cmsAuthorMark{1}, N.~Krammer, I.~Kr\"{a}tschmer, D.~Liko, T.~Madlener, I.~Mikulec, N.~Rad, J.~Schieck\cmsAuthorMark{1}, R.~Sch\"{o}fbeck, M.~Spanring, D.~Spitzbart, W.~Waltenberger, C.-E.~Wulz\cmsAuthorMark{1}, M.~Zarucki
% ...
\vskip\cmsinstskip
\dag: Deceased\\
1:  Also at Vienna University of Technology, Vienna, Austria\\
% ...
\end{sloppypar}
\end{document}